\documentclass[12pt]{article}
\pdfoutput=1%permet de dire que les figures sont des .pdf (pour JHEP)
\usepackage{graphicx,epstopdf,amssymb,amsfonts,amsmath,amsthm,array,
mathrsfs,amscd}%,wick}
\usepackage{todonotes} %remarques dans la marge avec la commande \todo
\newcommand{\pbs}[1]{\let\temp=\\#1\let\\=\temp}
\numberwithin{equation}{section}
\def\be{\begin{equation}}\def\ee{\end{equation}}
\def\cvp{\raise 2pt\hbox{,}}

 \def\d{{\rm d}}

\def\Im{\mathop{\rm Im}}
\def\Re{\mathop{\rm Re}}
\def\d{\partial}

\def\cM{{\cal M}}
\def\d{{\rm d}}

\def\a{\alpha}
\def\b{\beta}
\def\g{\gamma}
\def\G{\Gamma}
\def\dd{\delta}
\def\e{\epsilon}
\def\m{\mu}
\def\n{\nu}

\def\l{\lambda}

\def\s{\sigma}
\def\f{\phi}

\def\t{\theta}

\def\D{\Delta}

\def\vf{\varphi}
\def\L{\Lambda}

\def\wh{\widehat}
\def\wt{\widetilde}

\def\zb{\overline z}

\def\l{\lambda}

\def\ov{\overline}
\def\del{\partial}

\def\ba{\begin{eqnarray}}
\def\ea{\end{eqnarray}}
\theoremstyle{plain}% default
\theoremstyle{definition}
\theoremstyle{remark}

\def\imath#1#2#3{{\it Invent math }{\bf #1} (#2) #3}

\begin{document}
%
%\pagenumbering{roman}
%

%%%%%%%%%%%%%%%%%%%%%%%%%%%%%%%%%%%%%%%%%%%%%%%%%%%%%%%%%%%

{\pagestyle{empty}
\parskip 0in
\

\vfill
\begin{center}

{\Large \bf 2D gravitational Mabuchi action}

\medskip

{\Large \bf on Riemann surfaces with boundaries}

\vspace{0.2in}

Adel B{\scshape ilal} 
and Corinne de L{\scshape acroix}
\\
\medskip
\it {Laboratoire de Physique Th\'eorique de l'\'Ecole Normale Sup\'erieure\\
PSL Research University, CNRS, Sorbonne Universit\'es, UPMC Paris 6\\
24 rue Lhomond, F-75231 Paris Cedex 05, France}

\smallskip

%{\tt adel.bilal@lpt.ens.fr}
\end{center}
\vfill\noindent
We study  the gravitational action induced by coupling two-dimensional non-conformal, massive matter  to gravity on a Riemann surface with boundaries. A small-mass expansion gives back the Liouville action in the massless limit, while the first-order mass correction allows us to identify what should be the appropriate generalization of the Mabuchi action on a Riemann surface with boundaries. We provide a detailed study for the example of the cylinder. Contrary to the case of manifolds without boundary, we find that the gravitational Lagrangian explicitly depends on the space-point, via the geodesic distances to the boundaries, as well as on the modular parameter of the cylinder, through an elliptic $\t$-function.
%\vfill
%\smallskip
%
%\vfill
\begin{flushleft}
%\today
\end{flushleft}
%
%\newpage\pagestyle{plain}
\hrule

{\parskip -0.8mm
\small{\tableofcontents}}
}
\newpage
\setcounter{page}{1}

%%%%%%%%%%%%%%%%%%%%%%%%%%%%%%%%%%%%%%%%%%%%%%%%%%%%%%%%%%%

%%%%%%%%%%%%%%%%%%%%%%%%%%%%%%%%%%%%%%%%%%%%%%%%%%%%%%%%%%%
%
%
%%%%%%%%%%%%%%%%%%%%%%%%%%%%%%%%%%%%%%%%%%%%
\section{Introduction and generalities\label{intro}}

In two dimensions the standard Einstein-Hilbert action of gravity is a topological invariant and does not provide any dynamics for the metric. However, when matter is coupled to two-dimensional gravity with metric $g$ one may compute the matter partition function $Z_{\rm mat}[g]$ first and then define an ``effective" gravitational action as 
\be\label{Sgravgen}
S_{\rm grav}[g_0,g] =-\log \frac{Z_{\rm mat}[g]}{Z_{\rm mat}[g_0]}\ ,
\ee
where $g_0$ is some reference metric. This gravitational action then is to be used in the functional integral over the metrics, after appropriately fixing the diffeomorphism invariance.
Obviously any gravitational action defined this way will  satisfy a cocycle identity
\be\label{cocycle}
S_{\rm grav}[g_1,g_2] + S_{\rm grav}[g_2,g_3] =S_{\rm grav}[g_1,g_3]  \ .
\ee
Well-known examples of such gravitational actions are the Liouville \cite{Liouville1}, Mabuchi and Aubin-Yau actions \cite{Mabuchi,AubinYau}, as well as the cosmological constant action $S_c[g_0,g]=\m_0  \int \d^2 x (\sqrt{g} -\sqrt{g_0})=\m_0 (A-A_0)$.
While the Liouville action is formulated entirely in terms of $g_0$ and the conformal factor $\s$ (defined as $g=e^{2\s} g_0$), the Mabuchi and Aubin-Yau actions crucially involve  also directly the K\"ahler potential $\f$. In the mathematical literature they  appear in relation with the characterization of constant scalar curvature metrics.
Their roles as two-dimensional gravitational actions in the sense of \eqref{Sgravgen} have been put forward in \cite{FKZ}. In particuler, ref.~\cite{FKZ} has studied the metric dependence of the partition function of non-conformal massive matter on compact Riemann surfaces and shown that  a gravitational action defined by \eqref{Sgravgen} contains these Mabuchi and Aubin-Yau actions  as first-order corrections (first order in $m^2 A$ where $m$ is the mass and $A$ the area of the Riemann surface) to the Liouville action. The study of \cite{FKZ} was further confirmed and generalized in \cite{BLgrav} where an exact formula for the gravitational action for any value of the mass was obtained, its expansion in $m^2 A$ giving back the results of \cite{FKZ}. 

The Mabuchi action has drawn much attention in recent years. It has been suggested in \cite{FKZ} that it may serve as a candidate action for novel two-dimensional quantum gravity models, motivated partly by  applications in K\"ahler geometry \cite{FKZrandom}, such as its relation to the stability in K\"ahler geometry \cite{KZstab}. Some further physical properties of the Mabuchi action such as the critical exponent and the spectrum have been studied in \cite{BFK}, \cite{BL}, \cite{dLESletter} and \cite{dLESmini}.

Quite amazingly, the Mabuchi action also emerges as a subleading term in the gravitational effective action in Quantum Hall wave functions, such as the Laughlin state \cite{SKBergman} - \cite{KlargeN}. There the Mabuchi action corresponds to the Wen-Zee term in the Chern-Simons description of the Quantum Hall effect \cite{WZ} and the coefficient in front of it controls the celebrated Hall viscosity.

All these results and developments focussed on compact Riemann surfaces without boundaries. Obviously, it is most important to generalize the Mabuchi action, and more generally the determination of the subleading terms in the gravitational action, to the case where the Riemann surface has boundaries. In particular, with view on the relation to the Quantum Hall effect, obtaining the generalization of the effective action including the boundary effects would be most interesting.

Here, ou goal will be somewhat more modest. References \cite{FKZ}  and \cite{BLgrav} considered  a massive scalar field with action
\be\label{scalaraction}
S_{\rm mat}[g,X]=\frac{1}{2}\int_{\cM}\d^2 x \sqrt{g} \left[ g^{ab}\del_a X \del_bX + m^2 X^2\right]
=\frac{1}{2}\int_{\cM}\d^2 x \sqrt{g}\, X (\D_g + m^2)X  \ ,
\ee
living on a compact Riemann surface $\cM$ of genus $h$. As shown in these references, this
leads to a gravitational action that, when expanded in $m^2 A$ gives  the Liouville action
to lowest order, 
\be\label{Liouvaction}
S_L[g_0,g] \equiv S_L[g_0,\s]=\int_{\cM}\d^2 x \sqrt{g_0} \, \big( \s \D_0\s + R_0 \s\big) 
, \quad g=e^{2\s} g_0\ .
\ee
and,  to first order,  a combination of the Mabuchi and Aubin-Yau actions:
\ba\label{Mab1}
S_{\rm M}[g_0,g]
&=&\int_{\cM}\d^2 x \sqrt{g_0} \left[ 2\pi(h-1)\f\D_0\f + \Bigl(\frac{8\pi(1-h)}{A_0}-R_0\Bigr) \f +\frac{4}{A} \s e^{2\s} \right] \  ,\\
S_{\rm AY}[g_0,g]
&=&-\int_{\cM}\d^2 x \sqrt{g_0} \left[ \frac{1}{4} \f\D_0\f -\frac{\f}{A_0}\right] \ ,
\ea
where the K\"ahler potential $\f$ is related to the conformal factor $\s$ and the areas $A$  and $A_0$ of $\cM$ as measured by $g$ and $g_0$ through the relation
\be\label{gg0sig}
e^{2\s}= \frac{A}{A_0}\left(1-\frac{1}{2} A_0 \D_0\f\right)\ .
\ee

In this note, we want to study how these results get modified when the two-dimensional Riemann surface $\cM$ has boundaries $\del\cM$. A priori, two things could happen: the corresponding gravitational actions  could get additional  boundary contributions, and the bulk gravitational   Lagrangian at a point $x$ could explicitly depend on the geodesic distances between $x$ and the boundaries. We will indeed observe both of these.

Obviously, in the presence of boundaries, we have to impose some boundary conditions. Our choice will be guided by two requirements: we want $\D_g+m^2$, i.e. $\D_g=e^{-2\s}\D_{g_0}\equiv e^{-2\s}\D_{0}$ to be hermitian and we want to preserve the fact that $\int_{\cM}\d^2 x \sqrt{g}\, \D_g f \equiv \int_{\cM}\d^2 x \sqrt{g_0}\, \D_0 f = 0$. Recall that 
\be\label{Lapldef}
\D_g f=-\frac{1}{\sqrt{g}}\del_a ( \sqrt{g} g^{ab}\del_b f)\ .
\ee  
The hermiticity condition 
\be\label{Laplherm}
(\vf_1,\D_g \vf_2)\equiv\int_{\cM} \d^2 x \sqrt{g}\ \ov\vf_1 \D_g \vf_2
=\int _{\cM}\d^2 x \sqrt{g}\ \ov{\D_g\vf_1} \, \vf_2\equiv (\D_g\vf_1, \vf_2)
\ee
yields the vanishing of the boundary term
\be\label{Laplherm2}
\int_{\del\cM} \d l \,  n^a\, (\ov{\del_a \vf_1} \vf_2-\ov{\vf_1} \del_a \vf_2)\ ,
\ee
where $n^a$ is the normal vector of the boundary and $\d l$ the invariant line element on the boundary. (See the appendix for the definition of the normal vector). As usual, this leads to two possible choices of boundary conditions: either  $\vf=0$ (Dirichlet) or $n^a \del_a \vf=0$ (Neumann)  on the boundary. Actually, the modified Neumann (Robin) conditions  $n^a \del_a \vf= c \, \vf$ with real $c$ are also possible. Our second condition reads
\be\label{intdelta=0}
0=\int_{\cM}\d^2 x \sqrt{g}\, \D_g f=\int_{\del\cM}\d l\,  n^a \del_a f \ ,
\ee
selecting the Neumann boundary conditions. In particular, if the massive matter field(s) $X$ obey these boundary conditions, one may freely integrate by parts in the matter action and the equality of both expressions in \eqref{scalaraction} still holds for a manifold $\cM$ with boundaries.
From now on, we will always assume that the matter field(s) obey Neumann boundary conditions. What about the K\"ahler field $\f$ and the conformal factor $\s$~? It follow from \eqref{gg0sig} that $\f$  also must satisfy Neumann conditions (in the metric $g_0$). Indeed, the area should be given by $A=\int\sqrt{g_0}e^{2\s}$ which, by \eqref{gg0sig} implies that $0=\int\sqrt{g_0} \D_0\f$ which is possible only if $n_0^a \del_a\f=0$ on $\del \cM$. This same relation \eqref{gg0sig} also implies $\del_n\s=-\frac{A}{4}e^{-2\s} \del_n (\D_0\f)$, showing that it is not compatible to impose Neumann boundary conditions also on $\s$.

Our main result is the formula \eqref{Sgrav5} for the first-order (in $m^2 A$) correction to the gravitational action on a Riemann surface with boundaries:
\ba\label{Sgrav5intro}
S_{\rm grav}[g,g_0]&=&-\frac{1}{24\pi}  S_{L}[g,g_0] +\frac{1}{2}\log\frac{A}{A_0}
+\frac{ m^2\, (A-A_0)}{2 A_0}   \Phi_G[g_0] 
\nonumber\\
&&+ \frac{m^2 A}{4} \, \Bigg[ \int\sqrt{g_0}\Big(-\frac{1}{2}\f\D_0\f  +\frac{1}{\pi A} \s e^{2\s} -\D_0\f\,  \wt G^{(0)}_{\rm R,bulk}[g_0] \Big)  \Bigg]
+ {\cal O}(m^4)\ .
\ea
On the first line, $S_L$ is the Liouville action including the boundary contributions.
The second line explicitly shows the terms that generalize the Mabuchi action. Here $ \wt G^{(0)}_{\rm R,bulk}[g_0]$ is a certain renormalized Green's function ``at coinciding points" that depends on the point on the Riemann surface, and in particular on the geodesic distances to the various boundaries.  One could integrate by parts the Laplacian in $\D_0 \f \, \wt G^{(0)}_{\rm R,bulk}$ to get $ \f \, \D_0\wt G^{(0)}_{\rm R,bulk}$ instead, but this would also generate additional boundary terms since $ \wt G^{(0)}_{\rm R,bulk}$ does not obey the Neumann boundary conditions. We offer various re-writings of this expression involving different variations of renormalized Green's functions at coinciding points, see e.g. \eqref{Sgrav7}.

To get more insight into the meaning of this rather abstract formula, we worked out the simplest case of a Riemann surface with boundary, which is a cylinder. In this case the Green's function is well-known and we explicitly determined the various versions of renormalized Green's functions at coinciding points. As expected, these quantities  depend on the distance to the two boundaries of the cylinder. (In the case of the compact torus the corresponding functions are just constants.)
We explicitly determined them in terms of elliptic theta functions, cf \eqref{Sgravcyl1}~:
\be\label{Sgravcyl1intro}
S_{\rm grav}^{\rm cyl}[g,g_0]\Big\vert_{m^2 A{\rm -term}}
= \frac{m^2 A}{4} \,  \int_0^T \d x \int_0^{2\pi R} \d y\, \Big(-\frac{1}{2}\f\D_0\f  +\frac{1}{\pi A} \s e^{2\s} 
+\frac{1}{2\pi} \D_0\f\,  \log \t_1 \big(\frac{x}{T} \big\vert i\pi\frac{R}{T}\big)  \Big) \ .
\ee
Again, we offer some equivalent rewriting of this action, cf \eqref{Sgravcyl2}.

The plan of this paper is the following. In the next section we will set up the basic frame to compute the gravitational action from the appropriate spectral $\zeta$-function. The strategy is to determine the variation of the gravitational action under an infinitesimal change of the conformal factor of the metric and, in the end, to integrate this relation to obtain the gravitational action itself. In section 3, we introduce some standard tools -- Green's functions and heat kernels -- and discuss their specific features on manifolds with boundaries. In sect. 4 we show how these quantities are related to local $\zeta$-functions with emphasis and the singularities resulting from the boundaries. In sect. 5 we put everything together to determine the gravitational action. The lowest order term in a small mass expansion, of course, just gives back the Liouville action, including a boundary term, while the first-order term in $m^2 A$ gives us what we call the Mabuchi action on the manifold with boundaries, as written above in \eqref{Sgrav5intro}. Let us emphasize again that it does not involve a boundary term, but the bulk Lagrangian explicitly depends on the geodesic distances to the various boundary components. In section 6 we work out the explicit example of a cylinder and one sees this dependence through an elliptic $\t$-function that depends on the distances to the two boundaries and on the modular parameter of the cylinder, cf \eqref{Sgravcyl1intro}.    Finally, we study what happens for an infinitely long cylinder -- viewed as a model of euclidean time and space being a circle. In this case we find that the Mabuchi Lagrangian reduces to the standard Mabuci Lagrangian \eqref{Mab1} with $R_0=0$ and $h=0$.

%%%%%%%%%%%%%%%%%%%%%%%%%%%%%%%%%%%%%%%%%%%%

\section{The gravitational action\label{secgravaction}}

\subsection{Basics}

We call $\vf_n$ and $\l_n$ the  eigenfunctions
and eigenvalues of the hermitian (thanks to the boundary condition\footnote{
Obviously, one should not confuse the normal vector $n^a$ with the index $n$ labelling the eigenvalues and eigenfunctions.
}) differential operator appearing in $S_{\rm mat}$:
\be\label{eigenvv}
(\D_g+m^2)\vf_n=\l_n \vf_n \quad , \quad \int\d^2 x \sqrt{g}\, \vf_n \vf_m =\dd_{nm} 
\quad , \quad  n^a\del_a \vf_m = 0 \ {\rm on}\ \del\cM  \ .
\ee
Since $\D_g+m^2$ is real, one may choose the eigenfunctions $\vf_n$ to be real, which we will always assume (unless an obvious complex choice has been made, like the standard spherical harmonics on the round sphere).
We take the indices $n$ to be $n\ge 0$ with $n=0$ referring to the lowest eigenvalue. In particular,  the Laplace operator always has a constant zero-mode, $\vf_0=\frac{1}{\sqrt{A}}$ and thus $\l_0=m^2$, since this constant obviously  obeys the Neumann boundary condition. As usual, these $\vf_n$ form a complete set of eigenfunctions.

The matter partition function is defined  with respect to the decomposition  of the matter field on these eigenfunctions $\vf_n$ : $X=\sum_{n\ge 0} c_n \vf_n$  as
\be\label{Zmat}
Z_{\rm mat}[g]=\int{\cal D}_g X e^{-S_{\rm mat}[g,X]} 
= \int \prod_{n=0}^\infty \frac{\d c_n}{\sqrt{2\pi}} e^{-\frac{1}{2} \sum_{n\ge 0} \l_n c_n^2}
=\big(\det (\D_g+m^2) \big)^{-1/2} .
\ee
In the massless case, this has to be slightly modified, see \cite{FKZ,BLgrav}.

Of course, the determinant is ill-defined and needs to be regularized. We will use the very convenient regularization-renormalization in terms of the spectral $\zeta$-functions:
\be\label{zeta}
\zeta(s)=\sum_{n\ge 0} \l_n^{-s} \ ,
\ee 
By Weil's law (see e.g.~\cite{BF}), the asymptotic behaviour of the eigenvalues for large $n$ is $\l_n \sim \frac{n}{A}$ and, hence the spectral  $\zeta$-functions are defined by converging sums for ${\rm Re}\, s>1$, and by analytic continuations for all other values. In particular, they are well-defined meromorphic functions for all complex values of $s$ with a single pole at $s=1$ with residue $\frac{1}{4\pi}$ (see e.g. \cite{BF}). A straightforward formal manipulation shows that $\zeta'(0)\equiv \frac{\d}{\d s}\zeta(s)\vert_{s=0}$ provides a formal definition of $-\sum_{n\ge 0} \log \l_n$, i.e. of $-\log \det(\D_g+m^2)$:
\be\label{Zmatzeta}
Z_{\rm mat}[g]=\exp\Big({\frac{1}{2} \zeta'(0)}\Big) \ .
\ee

There is a slight subtlety one should take into account, see e.g. \cite{BF}. While the field $X$ is dimensionless, the $\vf_n$ scale as $A^{-1/2}\sim \wh\m$ where $\wh\m$ is some arbitrary mass scale (even if $m=0$), and the $c_n$ as $\wh\m^{-1}$. It follows that one should write ${\cal D}_g X=\prod_n \frac{\wh\m\d c_n}{2\pi}$. This results in $Z_{\rm mat}=\big(\prod_n \frac{\l_n}{\wh\m^2}\big)^{-1/2}$, so that  $\zeta'(0)$ is changed into
\be\label{zetamu}
\zeta'(0) \to \zeta'(0)+ \zeta(0)\, \log\wh\m^2 \ .
\ee

The regularization-renormalization of determinants in terms of the $\zeta$-function may appear as rather ad hoc, but it can be rigorously justified by introducing the spectral regularization \cite{BF}. The regularized logarithm of the determinant then equals $\zeta'(0)+\zeta(0)\log \wh\m^2$ plus a diverging piece $\sim A\L^2 (\log\frac{\L^2}{\wh\m^2} + {\rm const})$, where $\L$ is some cutoff. This diverging piece just contributes to the cosmological constant action, and this is why the latter must be present as a counterterm, to cancel this divergence.
Thus, we finally arrive at
\be\label{Sgravgen2}
\hskip-1.cm S_{\rm grav}[g_0,g]=-\frac{1}{2} \left(\zeta_g'(0)+\zeta_g(0) \log\wh\m^2\right) +\frac{1}{2} \left(\zeta_{g_0}'(0)  +\zeta_{g_0}(0)\log\wh\m^2\right)\, .
\ee
It is important to notice that this formula expressing the gravitational action in terms of the $\zeta$-function is true whether the Riemann surface has a boundary or not. Of course, the $\zeta$-function for a manifold with boundary will have some properties that differ from the case without boundary. Formally, the $\zeta$-functions are always defined by \eqref{zeta}, but the properties of the manifold are encoded in the eigenvalues $\l_n$ that appear in the sum.

The strategy of \cite{FKZ} and \cite{BLgrav}, that we will also follow here, was to determine the infinitesimal change of the $\zeta$-functions from the infinitesimal change of the eigenvalues $\l_n$ under an infinitesimal change of the metric, and then to integrate this relation to get $S_{\rm grav}$.
The change of the eigenvalues is obtained from (almost) standard quantum mechanical perturbation theory, as we discuss next.

\subsection{Perturbation theory}

We want to study how the eigenvalues $\l_n$ and eigenfunctions $\vf_n$ change under an infinitesimal change of the metric. Since $g=e^{2\s}g_0$, the Laplace operator $\D_g$ and hence also $\D_g+m^2$ only depend on the conformal factor $\s$ and on $g_0$:
$\D_g=e^{-2\s}\D_{0}$ and thus under a variation $\dd\s$ of $\s$ one has
\be\label{Lapvar}
\dd \D_g=-2\dd\s \D_g \quad \Rightarrow\quad
\langle \vf_k | \dd \D_g | \vf_n\rangle = -2(\l_n-m^2) \langle \vf_k | \dd \s | \vf_n\rangle \ ,
\ee
where, of course, $\langle \vf_k | \dd \s | \vf_n\rangle=\int\d^2 x \sqrt{g}\, \vf_k \dd\s \vf_n$. 
One can then apply standard quantum mechanical perturbation theory. The only subtlety comes from the normalisation condition in \eqref{eigenvv} which also gets modified when varying $\s$ \cite{FKZ,BF}. One finds
\ba\label{deltalambda}
\dd\l_n&=&-2(\l_n-m^2) \langle \vf_n | \dd \s | \vf_n\rangle \ ,
\\
\label{deltaphi}
\dd\vf_n&=&- \langle \vf_n | \dd \s | \vf_n\rangle\, \vf_n - 2 \sum_{k\ne n} \frac{\l_n-m^2}{\l_n-\l_k} \langle \vf_k | \dd \s | \vf_n\rangle \, \vf_k \ .
\ea
Let us insists that this is first-order perturbation theory in $\dd\s$, but it is exact in $m^2$.
Note the trivial fact that, since $\l_0=m^2$, one consistently has 
\be\label{deltalamzero}
\dd\lambda_0=0 \ .
\ee

\subsection{Variation of the determinant}

As mentioned above, in order to compute  $S_{\rm grav}[g_0,g]$ as given by  \eqref{Sgravgen2}, we will compute $\dd \zeta'(0)\equiv \dd \zeta_g'(0)$ and $\dd \zeta(0)\equiv \dd \zeta_g(0)$ and express them as  ``exact differentials" so that one can integrate them and obtain the finite differences $\zeta_{g_2}'(0)-\zeta_{g_1}'(0)$ and $\zeta_{g_2}(0)-\zeta_{g_1}(0)$.

From \eqref{deltalambda} one immediately gets, to first order in $\dd\s$,
\be\label{deltazeta}
\zeta_{g+\dd g}(s)=\sum_{n\ge 0}\frac{1}{(\l_n+\dd\l_n)^s}
=\zeta_g(s) + 2 s \sum_{n\ge 0} \frac{\l_n-m^2}{\l_n^{s+1}}  \langle \vf_n | \dd \s | \vf_n\rangle \ ,
\ee
As noted before, $\dd\l_0=0$ and, hence, there is no zero-mode contribution to the second term and one could just equally well rewrite the following results in terms of the $\wt\zeta$-functions defined by excluding the zero-mode \cite{BLgrav}. Here, however, this is not of particular interest to us, and thus
\be\label{deltazeta2}
\dd\zeta(s)=2 s \int\d^2 x\sqrt{g}\  \dd\s(x) \big[ \zeta(s,x,x)-m^2 \zeta(s+1,x,x) \big] \ ,
\ee
where
\be\label{localzetadef}
\zeta(s,x,y)=\sum_{n\ge 0} \l_n^{-s}\, \vf_n(x)\vf_n(y) 
\ee
is a (bi)local $\zeta$-function.
As we will see, $\zeta(s,x,x)$ has a pole at $s=1$ for every $x$. For $x$ in the bulk, this pole is the only singularity. However, as $x$ goes to the boundary there could be, a priori, additional singularities for other values of $s$, in particular for $s=0$. Keeping this in mind we find 
\ba\label{deltazeta3}
\dd\zeta'(0)&=&2 \Bigg\{ \lim_{s\to 0} \Big[ 1+s\frac{\d}{\d s} \Big] 
- m^2\,  \lim_{s\to 1} \Big[ 1+(s-1) \frac{\d}{\d s} \Big] \Bigg\} \, \int\d^2 x\sqrt{g}\,  \dd\s(x)  \zeta(s,x,x) 
\nonumber\\
\dd\zeta(0)&=&2 \Bigg\{ \lim_{s\to 0} \ s  \ \ 
-m^2 \, \lim_{s\to 1}\ (s-1) \ \Bigg\} \, \int\d^2 x\sqrt{g}\  \dd\s(x)  \zeta(s,x,x)
\ .
\ea
The rest of this paper is devoted to computing $\int\d^2 x\sqrt{g}\  \dd\s(x)  \zeta(s,x,x)$ on a Riemann surface with boundaries and extracting its behaviour as $s\to 0$ and $s\to 1$. More generally, we will compute $\int\sqrt{g} f(x) \zeta(s,x,x)$ where $f$ is some sort of ``test function". Once we have determined these quantities, we will get the variation of the gravitational action under an infinitesimal change of metric as
\ba\label{deltaSgrav}
\dd S_{\rm grav}&=&-\frac{1}{2} \dd \zeta'(0) -\frac{1}{2} \dd\zeta(0)\, \log\wh\m^2 \nonumber\\
&&\hskip-1.5cm = -\Bigg\{ \lim_{s\to 0} \Big[ 1+s\big(\frac{\d}{\d s} +  \log\wh\m^2\big)\Big] 
- m^2\,  \lim_{s\to 1} \Big[ 1+(s-1)\big( \frac{\d}{\d s}+ \log\wh\m^2 \big)\Big] \Bigg\} \int\d^2 x\sqrt{g}\,  \dd\s(x) \zeta(s,x,x) 
 \ .
 \nonumber\\
\ea
Finally note that the variations of the conformal factor $\dd\s$, of the K\"ahler potential $\dd\f$ and  the area $\dd A$ are related as
\be\label{delsigmadelphirel}
\dd\s=\frac{\dd A}{2 A} -\frac{A}{4} \D \dd\f \ .
\ee

%\newpage
%%%%%%%%%%%%%%%%%%%%%%%%%%%%%%%%%%%%%%%%%%%%
\section{Some technical tools : Green's functions and the heat kernel}

In this section we discuss some standard technical tools. We assume that the Riemann surface $\cM$ has a boundary $\del\cM$ and that we have imposed Neumann boundary conditions. Throughout this section we assume that some fixed metric $g$ has been chosen on $\cM$.

\subsection{Complete set of eigenfunctions and Green's functions}

Recall from \eqref{eigenvv} that the $\vf_n$ and  $\l_n$ are the orthonormal eigenfunctions and eigenvalues of\break\hfill $(\D_g+m^2)$, subject to the Neumann boundary condition $n^a\del_a \vf_m = 0 \ {\rm on}\ \del\cM$.
They form a complete set which means that they obey the
completeness relation 
\be\label{complete}
\sum_{n\ge 0} \vf_n(x) \vf_n(y) =\frac{1}{\sqrt{g}} \dd(x-y) \ , 
\quad x,y\in \cM \setminus \del\cM \ .
\ee
Actually, this continues  to hold also  if $x$ or $y$ are on the boundary\footnote{
Suppose $y=y_B$ is on the boundary. Then the $\dd(x-y_B)$ integrates to $\frac{1}{2}$ rather than 1. On the other hand, as will become clear below, in this case one actually has two $\dd$'s, the factor 2 off-setting the factor $\frac{1}{2}$. With this understanding, \eqref{complete} continues to hold also on the boundary. 
} 

As always, the Green's function of an operator like $\D_g+m^2$ can be given in terms of the eigenfuctions and eigenvalues as
\be\label{Greeneigen}
G(x,y)=\sum_{n\ge 0} \frac{\vf_n(x)\vf_n(y)}{\l_n} \ , \quad 
(\D_g+m^2) G(x,y) =\frac{1}{\sqrt{g}}\dd(x-y) \ , \quad x,y\in \cM \setminus \del\cM \ .
\ee
Again, this continues  to hold also  if $x$ or $y$ are on the boundary.
In the massless case, $\l_0=0$ and this zero-mode must be excluded from the sum. We put a tilde on all quantities from which the zero-mode has been excluded:
\be\label{Greeneigenmassless}
\wt G(x,y)=\sum_{n> 0} \frac{\vf_n(x)\vf_n(y)}{\l_n} \ , \quad 
(\D_g+m^2) \wt G(x,y) =\frac{1}{\sqrt{g}}\dd(x-y) -\frac{1}{A}\ , \quad x,y\in \cM \setminus \del\cM \ .
\ee
Furthermore, we will add a superscript $(0)$ on all quantities that refer to the massless case. Obviously, $G(x,y)$, $\wt G(x,y)$ and $\wt G^{(0)}(x,y)$ satisfy the Neumann boundary conditions in each of their arguments.

\subsection{The heat kernel}
The heat kernel and integrated heat kernel for the operator $\D_g+m^2$ are similarly defined in terms of the eigenvalues and eigenfunctions \eqref{eigenvv} as
\be\label{heatkernel}
K(t,x,y)=\sum_{n\ge0} e^{-\l_n t} \,\vf_n(x) \vf_n(y) 
\quad , \quad
K(t)=\int\d^2 x \sqrt{g}\, K(t,x,x)=\sum_{n\ge 0} e^{-\l_n t} \ .
\ee
%The corresponding $\wt K$, $K^{(0)}$ and $\wt K^{(0)}$ are defined analogously, where the tilde  means to exclude the ``zero-mode" and the superscript $(0)$ refers to the same quantities for $m=0$. 
It is obvious from this definition that $K(t,x,y)$ satisfies the Neumann boundary conditions in both arguments $x$ and $y$ and is the solution of
\be\label{heatdiffeq}
\left(\frac{\d}{\d t} + \D_g+m^2\right) K(t,x,y) = 0 
\quad , \quad
K(t,x,y)\sim\frac{1}{\sqrt{g}} \dd(x-y)\ \text{as\ } t\to 0 \ .
\ee
%The last condition holds for all $x$ and $y$ in the interior of $\cM$, while for $x$ or $y$ in $\del\cM$, the same remarks as above apply.
Note that it immediately follows from either \eqref{heatkernel} or \eqref{heatdiffeq} that the massless and massive heat kernels are simply related by
%\be\label{KmKo}
$K(t,x,y)=e^{-m^2 t} K^{(0)}(t,x,y)$.
%\ee
As is also clear from  \eqref{heatkernel}, for $t>0$, $K(t,x,y)$ is given by a converging sum and is finite, even as $x\to y$. For $t\to 0$ one recovers various divergences, and, in particular 
\be\label{KG}
\int_0^\infty \d t \, K(t,x,y) = G(x,y)
\ee
exhibits the short distance singularity of the Green's function which is well-known to be logarithmic.

The behaviour of $K$ for small $t$ is related to the asymptotics of the eigenvalues $\l_n$ and eigenfunctions $\vf_n$ for large $n$, which in turn is related to the short-distance properties of the Riemann surface. It is thus not surprising that the small-$t$ asymptotics is given in terms of local expressions of the curvature and its derivatives. Indeed, {\it on a compact manifold without boundaries} one has the well-known small $t$-expansion\footnote{we write ${\cal K}$ for the heat kernel on a compact manifold without boundaries}:
\be\label{Kasymp}
{\cal K}(t,x,y) \sim \frac{1}{4\pi t} e^{- m^2 t -\ell^2(x,y)/4t } \sum_{k\ge 0} t^k a_k(x,y)\ ,
\ee
where $\ell^2(x,y)\equiv\ell^2_g(x,y)$ is the geodesic distance squared between $x$ and $y$. For small $t$, the exponential forces $\ell^2$ to be small (of order $t$) and one can use normal coordinates around $y$. This allows one to obtain quite easily explicit expressions for the $a_r(x,y)$ in terms of the curvature tensor and its derivatives. They can be found e.g.~in \cite{BF} and, in particular, $a_0(x,x)=1$ and $a_1(x,x)=\frac{R(x)}{6}$.  At present, on a manifold $\cM$ with boundaries, this asymptotic expansion must still be valid as long as $x$ and $y$ are  ``not too close" to the boundary. However, as the points get close to the boundary, we expect extra contributions to become important.

\subsection{Examples}

Before going on, it is useful to discuss some very simple examples of manifolds with boundaries: the one-dimensional interval, the two-dimensional cylinder which is the product of the interval and a circle, and  the two-dimensional half-sphere.

\subsubsection{Example 1: the one-dimensional interval\label{ex1}}

The  simplest example is a one-dimensional manifold that is just the interval $\cM=[0,\pi]$ with trivial metric and $\D=-\del_x^2$. We also take $m=0$. The normalized eigenfunctions that satisfy the Neumann boundary conditions are $\vf_0=\frac{1}{\sqrt{\pi}}$ and $\vf_n(x)=\sqrt{\frac{2}{\pi}} \cos n x$, $n=1,2,\ldots$, and the eigenvalues are $\l_n=n^2$. Then we formally have for any function $f(\l)$:
\be\label{intervalsum}
\sum_{n=0}^\infty f(n^2 )\vf_n(x) \vf_n(y)
=\frac{1}{\pi} \Big[ f(0)+\sum_{n=1}^\infty f(n^2) 2\cos nx\, \cos ny\Big]
%\nonumber\\
%&&=\frac{1}{\pi} \Big[ f(0)+\sum_{n=1}^\infty f(n^2) \big( \cos n(x-y) + \cos n(x+y)\big)\Big]
=\frac{1}{2\pi}\sum_{n=-\infty}^\infty f(n^2) \Big[e^{in(x-y)} +e^{in(x+y)} \Big]
\ee
For $f=1$ this is just the completeness relation \eqref{complete} with the right-hand side equal to $\dd(x-y)+\dd(x+y)$ where  the $\dd$ are $2\pi$-periodic Dirac distributions, i.e. defined on the circle $S^1=[0,2\pi]$. With $x,y\in \cM \setminus \del\cM =]0,\pi[$, $x+y$ never is $0\mod 2\pi$ and the $\dd(x+y)$ never contributes. Actually, $x+y=x-y_C$, where $y_C=-y$ is the image point of $y$ due to the boundary at $y=0$. One would also expect additional image points due to the second boundary, but because of the $2\pi$ periodicity, these additional image points are equivalent to $y$ and $y_C$. If $y=y_B$  is on the boundary, say $y=0$, then the image point $y_C$ coincides with $y$ (possibly$\mod 2\pi$) and we get $2\dd(x-y_B)=2\dd(x)$. But $\int_0^\pi \d x \dd(x)=\frac{1}{2}$ and in any case the integral of the right-hand side of \eqref{complete} correctly gives 1.

If we let $f(n^2)=\frac{1}{n^2}$, the relation \eqref{intervalsum} expresses the Green's function $G_I$ on the interval with Neumann boundary conditions in terms of a sum of Green's functions ${\cal G}_{S^1}$ on the circle:\footnote{
Again, we notationally distinguish the Green's function on the compact manifold without boundaries, denoted as ${\cal G}$ from the one on the manifold with boundary, denoted as $G$.
}
\be\label{Greeninterval}
G_I(x,y)={\cal G}_{S^1}(x,y)+ {\cal G}_{S^1}(x,-y) \ ,
\ee
a construction well-known as the method of images. 

Similarly, if we let $f(n^2)=e^{-t n^2}$ we get a relation that expresses the heat kernel on the interval $K_I(t,x,y)$ as the sum of two heat kernels ${\cal K}_{S^1}$ on the circle, one at $x,y$ and the other at $x,-y$:
\be\label{heatinterval}
K_I(t,x,y)={\cal K}_{S^1}(t,x,y)+{\cal K}_{S^1}(t,x,-y) \ .
\ee
Actually, the sums can be expressed in terms of the theta function
\be\label{theta3def}
\t_3(\n | \tau)=\sum_{n=-\infty}^\infty q^{n^2} e^{2\pi i n \n} \quad , \quad q=e^{i\pi \tau}
\ee 
as
\be\label{Ktheta}
{\cal K}_{S^1}(t,x,y)=\frac{1}{2\pi}\, \t_3\big( \frac{x-y}{2\pi} \big\vert i\frac{t}{\pi}\big) \ .
\ee
The small-$t$ asymptotics is obtained by applying Poisson resummation, or equivalently the modular transformation of $\t_3$ under $\tau\to-\frac{1}{\tau}$,
\be\label{theta3modular}
\t_3(\n | \tau)=(-i\tau)^{-1/2} e^{-i\pi \n^2/\tau} \t_3\big( 
\frac{\n}{\tau} \big\vert -\frac{1}{\tau}\big) 
\ ,
\ee
which yields 
\be\label{Kcompexp}
{\cal K}_{S^1}(t,x,y) =\frac{1}{\sqrt{4\pi t}} \sum_{n=-\infty}^\infty e^{-(x-y+2\pi n)^2/(4t)} \ ,
\ee
and which expresses the heat kernel on the circle as a sum over all geodesics going from $x$ to $y$, winding an arbitrary number $n$ times around the circle and having length squared $(x-y+2\pi n)^2$. This is of course the expected result for the diffusion (Brownian motion) on a circle.
For small $t$, the leading term in ${\cal K}_{S^1}(t,x,y)$ always is the $n=0$ term. For ${\cal K}_{S^1}(t,x,-y)$, however, the leading term is $n=0$ if $x+y<\pi$, while it is $n=-1$ if $\pi<x+y$. Thus
%$\simeq \frac{1}{\sqrt{4\pi t}} e^{-(x\mp y)^2/(4t)} \big( 1+2e^{-\pi^2/t} \cosh \frac{\pi(x\mp y)}{t} +\ldots \big)$. Note that this series representation is valid only for $|x\mp y|<\pi$. In particular, we may encounter $\pi\le x+y$. In this case one must first use the $2\pi$ periodicity of  \eqref{Ktheta} and only then perform the transformation \eqref{theta3modular}.   This yields the following small-$t$ asymptotics\footnote{
%As $x\mp y\to \pi$ one has $e^{-\pi^2/t} \cosh \frac{\pi(x\mp y)}{t} \to 1$ and this is why the series representation ceases to converge. However, for $x\mp y\simeq \pi$ the prefactor $e^{-(x\mp y)^2/(4t)}$ is $e^{-\pi^2/(4t)}$ and is exponentially small for small $t$. It is thus justified to drop the terms $+2e^{-\pi^2/t} \cosh \frac{\pi(x\mp y)}{t} +\ldots$ since they either vanish exponentially for small $t$ or, if they don't, the whole expression is exponentially small.
%}
\ba\label{heatintas}
K_I(t,x,y)&\simeq\atop{t\to 0}& \frac{e^{-(x-y)^2/(4t)}}{\sqrt{4\pi t}} 
+ \frac{e^{-(x+y)^2/(4t)} }{\sqrt{4\pi t}} \  , \quad {\rm for}\ 0\le x+y\le\pi \ ,
\nonumber\\
K_I(t,x,y)&\simeq\atop{t\to 0}& \frac{e^{-(x-y)^2/(4t)}}{\sqrt{4\pi t}} 
+ \frac{e^{-(2\pi-x-y)^2/(4t)} }{\sqrt{4\pi t}} \  , \quad {\rm for}\ \pi\le x+y\le 2\pi \ .
\ea
Of course, only the coefficient $\frac{a_0}{\sqrt{4\pi}}=\frac{1}{\sqrt{4\pi}}$ appears, since all other $a_r$ involve the curvature and vanish in our present example. In any case, we see that for small $t$, the first term is exponentially small unless $x$ is close to $y$ within a distance of order $\sqrt{t}$. Similarly, the second term is exponentially small unless  $x+y$ (or $2\pi-x-y$) is of order $\sqrt{t}$ which is possible only  if $x$ and $y$ both are close to the boundary at 0 (or at $\pi$), and thus also close to each other, within a distance of order $\sqrt{t}$. Thus, for $x$ or $y$ in the bulk, the second term does not contribute to the small-$t$ expansion. It is only if both points go to one and the same boundary that the second term becomes important. We see that we can just as well write this small-$t$ asymptotic expansion as
\ba\label{heatintas2}
K_I(t,x,y)&\simeq\atop{t\to 0}& \frac{e^{-(x-y)^2/(4t)}}{\sqrt{4\pi t}} 
+\sum_{\del \cM_i}\  \frac{e^{-(x-y_C^{(i)})^2/(4t)} }{\sqrt{4\pi t}} \ ,
\ea
where the sum is over the different boundary components and $y_C^{(i)}$ denotes the image (``conjugate") point of $y$ with respect to the boundary component $\del \cM_i$, i.e.~ $y_C^{(1)}=-y$ and $y_C^{(2)}=2\pi-y$. 
While the use of image points is familiar from solving the Laplace equation for simple geometries in the presence of boundaries, we have seen that we should actually think of $(x-y_C^{(i)})^2$ as the length squared of the geodesic from $x$ to $y$ that is reflected once at the boundary $\del\cM_i$.  Geodesics with multiple reflections necessarily are much longer and give exponentially subleading contributions. Of course, if one uses the exact expression \eqref{Kcompexp} for ${\cal K}_{S^1}(t,x,y)+ {\cal K}_{S^1}(t,x,-y)$ the heat kernel of the interval is expressed as a sum over all geodesic paths from $x$ to $y$  being reflected an arbitrary number of times at the two boundaries.

%%%%%%%%%%%%%%%%%%%%%%
\subsubsection{Example 2: the cylinder\label{cylexample}}

The two-dimensional cylinder is just an interval times a circle, $I\times S^1$. Thus, if we choose the interval of length $a$ and the circle of circumference $2b$, the normalized eigenfunctions of the Laplace operator satisfying the Neumann boundary conditions are
\be\label{cyleigenfct}
\vf_{0,m}(x_1,x_2)=\frac{e^{i\pi m x_2/b}}{\sqrt{2ab}} \quad , \quad
\vf_{n,m}(x_1,x_2)=\frac{e^{i\pi m x_2/b}}{\sqrt{ab}} \, \cos \frac{n\pi x_1}{a} \quad , \quad m\in {\bf Z} \ , \ n=1,2,\ldots \ .
\ee
The heat kernel for the Laplace operator then simply is the product of the heat kernel for the circle and the heat kernel of the interval as just given in the previous example, with the obvious replacements $\pi\to a,b$~:
\ba\label{cylK}
K_{\rm cyl}(t,x_1,x_2,y_1,y_2)&=& \sum_{m=-\infty}^\infty \sum_{n=0}^\infty \exp{\Big( -t\big(\frac{\pi^2 n^2}{a^2}+\frac{\pi^2  m^2}{b^2}\big)\Big)}\, \vf_{n,m}(x_1,x_2)\, \vf_{n,m}^*(y_1,y_2)
\nonumber\\
&=&\frac{1}{4ab} \, \t_3\big(  \frac{x_2-y_2}{2b} \big\vert  i\frac{\pi t}{b^2}\big)
\Big[  \t_3\big( \frac{x_1-y_1}{2a}  \big\vert   i\frac{\pi t}{a^2}\big) 
+ \t_3\big(\frac{x_1+y_1}{2a}  \big\vert   i\frac{\pi t}{a^2}\big)
\Big]
\nonumber\\
\nonumber\\
&=& {\cal K}_{\rm torus}(t,x_1,x_2,y_1,y_2) + {\cal K}_{\rm torus}(t,x_1,x_2,-y_1,y_2)
 \ ,
\ea
with the corresponding torus obviously having periods $2a$ and $2b$. Poisson resummation or equivalently the modular transformation formula for $\t_3$ yields
\be\label{cylKresummed}
K_{\rm cyl}(t,x_1,x_2,y_1,y_2)=\frac{1}{4\pi t} \sum_{n,m=-\infty}^\infty \exp{\Big( -\frac{(x_1-y_1+2na)^2 +(x_2-y_2+2mb)^2}{4t}\Big)} \ + \  (y_1\to -y_1) \ .
\ee
Again, this expresses the heat kernel as a sum over all geodesics going from $x$ to $y$ winding $m$ times around the circle direction of the cylinder and being reflected $2n$ times (for the first term) or $2n+1$ times (for the second term) at the boundaries of the cylinder.

\subsubsection{Example 3: The upper half sphere}

Our last example involves a curved two-dimensional manifold with a boundary: let  $\cM$ be the upper half of the standard round sphere of unit radius, i.e. $\cM=S^2_+$, parametrized by $\t\in[0,\frac{\pi}{2}]$ and $\f\in[0,2\pi]$. Then the boundary $\del\cM$ is just the circle at $\t=\frac{\pi}{2}$ and the normal derivative is $n^a\del_a=\del_\t$. The eigenfunctions of the Laplace operator $\D$ on the sphere $S^2$ are the spherical harmonics $Y_l^m$ and, obviously, they still satisfy $\D Y_l^m=l(l+1)Y_l^m$ on $S^2_+$. However, not all of them satisfy the Neumann boundary condition.
As is well known, the parity of the $Y_l^m$ is $(-)^l$, so that 
\be\label{Ylmparity}
Y_l^m(\t,\f)=(-)^l Y_l^m(\pi-\t,\f+\pi)=(-)^{l-m} Y_l^m(\pi-\t,\f) \ .
\ee 
It follows that $Y_l^m$ is even (odd) under reflection by the equator at $\t=\frac{\pi}{2}$ if $l-m$ is even (odd), and hence satisfies Neumann (Dirichlet) conditions at $\t=\frac{\pi}{2}$.
Thus, for each $l$, there are $l+1$ allowed values of $m$.   It follows for even $l-m$ that $\int_{S^2_+} \ov{Y_{l_1}^{m_1}} Y_{l_2}^{m_2}=\frac{1}{2}\int_{S^2} \ov{Y_{l_1}^{m_1}} Y_{l_2}^{m_2}=\frac{1}{2}\dd_{l_1 l_2}\dd_{m_1 m_2}$.  We see that the orthonormal eigenfunctions $\vf_n$ of the Laplace operator on $\cM$ obeying the boundary conditions simply are the $\sqrt{2} Y_l^m$ with $l-m$ even. It also follows from \eqref{Ylmparity} that $Y_l^m(\t,\f)+Y_l^m(\pi-\t,\f)$ vanishes for odd $l-m$ and equals $2Y_l^m(\t,\f)$ for even $l-m$. Thus we have for any function $f(\l)$ the formal relation 
\ba\label{S2+S2}
&&\sum_{l,m\atop l-m {\ \rm even}}  f\big(l(l+1)\big) \ov{\sqrt{2}Y_l^m(\t,\f)} \sqrt{2} Y_l^m(\t',\f') 
\nonumber\\
&&\hskip2.cm =\sum_{l,m}  f\big(l(l+1)\big) \ov{Y_l^m(\t,\f)} \Big( Y_l^m(\t',\f')+Y_l^m(\pi-\t',\f')\Big) \ ,
\ea 
where it is of course understood that $l\ge 0$ and $|m|\le l$. If we simply take $f=1$, this is just the completeness relation, with it's right-hand side being
\be
\dd(\cos\t-\cos\t')\dd(\f-\f') + \dd(\cos\t+\cos\t')\dd(\f-\f') \ .
\ee
Here, the first term is just $\frac{1}{\sqrt{g}}\dd(x-y)$ while the second $\dd$ is $\frac{1}{\sqrt{g}}\dd(x-y_C)$, where $y_C=(\pi-\t',\f')$ is the ``image point" of $y=(\t',\f')$. As for the interval and the cylinder, this image point  is always outside of $\cM$, except if $y$ is on the boundary.  In the latter case both $\dd$'s contribute equally and one has $2\dd(\cos\t)\dd(\f-\f')$ which correctly gives 1 when integrated over $S^2_+$.
If necessary, this shows again that the completeness relation \eqref{complete} continues to hold for $x$ or $y$ on the boundary.

If we let $f\big(l(l+1)\big)=\frac{1}{l(l+1)+M^2}$ in \eqref{S2+S2} this relation expresses the Green's functions of $\D+M^2$ on the upper half sphere $S^2_+$ in terms of a sum of two Green's function on the sphere, one at $x$ and $y$ and the other at $x$ and $y_C$:
\be\label{upperspheregreen}
G_{S^2_+}(\t,\f;\t',\f')=
{\cal G}_{S^2}(\t,\f;\t',\f')+{\cal G}_{S^2}(\t,\f;\pi-\t',\f') \ .
\ee
An analogous relation holds for the $\wt G$ when the zero-mode is excluded, as well as for $\wt G^{(0)}$ when $M=0$ and the zero-mode is excluded.
It is interesting to study the short-distance singularity of this Green's function. In two dimensions, the short-distance singularity of the Green's function is logarithmic, and one has e.g. $\wt G^{(0)}_{S^2}(\t,\f;\t',\f)\sim -\frac{1}{4\pi} \log (\t-\t')^2$ as $\t\to\t'$. Then, on the half sphere, the singularity as $\t\to\t'$ for any $(\t',\f')\notin \del\cM$ is given by this same logarithmic singularity, since  $\wt G^{(0)}_{S^2}(\t,\f;\pi-\t',\f)$ is non-singular. However, if $(\t',\f')\in\del\cM$, i.e. $\t'=\frac{\pi}{2}$, then the short-distance singularity of $\wt G^{(0)}_{S^2_+}$ is twice as large, i.e. $-\frac{1}{2\pi}\log (\t-\t')^2$, in agreement with the factor 2 that accompanied the $\dd(x-y_B)$.

Finally, taking $f(\l)=e^{-t \l}$, we get the corresponding relation between the heat kernels:\break
$K_{S^2_+}(t,\t,\f;\t',\f')= {\cal K}_{S^2}(t,\t,\f;\t',\f')+{\cal K}_{S^2}(t,\t,\f;\pi-\t',\f')$.

%%%%%%%%%%%%%%%%%%
\subsection{The heat kernel continued}

As it appeared from the previous examples, in simple geometries, the Green's functions and the heat kernel can be obtained by a method of images from the corresponding Green's functions or heat kernels on a ``bigger" manifold without boundary, by a method of images.  In all three cases we have seen that
\be\label{ImageK}
K(t,x,y)={\cal K}(t,x,y)+{\cal K}(t,x,y_C) \ ,
\ee 
where ${\cal K}$ is the heat kernel on the ``bigger" compact manifold and $y_C$ the ``image point" of $y$. However, we have also seen in the example of the interval that the leading term in the asymptotic small-$t$ expansion to be used for  ${\cal K}(t,x,y_C)$ differs depending on whether $x$ and $y$ are close to one or the other boundary. 
Thus the small-$t$ asymtotic expansion has the following form
\be\label{Kasympsum}
K(t,x,y)\sim \frac{e^
{-m^2 t}}{4\pi t} \Big[ e^{-\ell^2(x,y)/4t} \sum_{k\ge 0} t^k a_k(x,y)+  \sum_{\del\cM_i}
 e^{-\ell^2(x,y_C^{(i)})/4t} \sum_{k\ge 0} t^k \wt a_k^{(i)}(x,y_C^{(i)}) \Big]  \ .
 \ee
Indeed, for $m=0$, the heat kernel describes the diffusion (Brownian motion) of a particle on the manifold from $x$ to $y$. On a flat manifold, this is given as a sum over the geodesic paths from $x$ to $y$ as $\frac{1}{4\pi t} e^{-\ell^2_{(r)}(x,y)/(4t)}$, where $\ell_{(r)}(x,y)$ is the geodesic length of the $r^{\rm th}$ path. In particular, if the manifold has boundaries, there are (possibly infinitely) many geodesic paths that involve one or several reflections at the boundaries. We write $\ell_{i}(x,y)$ for the length of the  geodesic path from $x$ to $y$ that involves exactly one reflection  at the boundary component $\del\cM_i$. Moreover, on a curved manifold, each $\frac{1}{4\pi t} e^{-\ell^2_{(r)}(x,y)/(4t)}$ gets multiplied by a power series in $t$ with coefficients that can be determined order by order from the differential equation \eqref{heatdiffeq}, see e.g.~\cite{Osborn}. For small $t$ the leading terms can involve at most one reflection, resulting indeed in the form \eqref{Kasympsum}, with $\ell^2(x,y_C^{(i)})$ replaced by  $e^{-\ell^2_i(x,y)/4t}$.   However, for small $t$, the terms involving $e^{-\ell^2_i(x,y)/4t}$ with one reflection at $\del\cM_i$ can only contribute if the points $x$ and $y$ are close to the boundary $\del\cM_i$, and close to each other, within a distance $\simeq\sqrt{t}$. As $t\to 0$, one zooms in close to the boundary which thus becomes flat. Now for a flat boundary, the length of the geodesic from $x$ to $y$ involving one reflection at $\del\cM_i$ is the same as the length of the geodesic from $x$ to the ``mirror" image point $y_C^{(i)}$ and $e^{-\ell^2_1(x,y)/4t}\simeq e^{-\ell^2(x,y_C^{(i)})/4t}$, with any differences at finite $t$ being included in a redefinition of the coefficients $\wt a_k^{(i)}(x,y_C^{(i)})\to a_k^{(i)}(x,y)$. Thus, we can rewrite \eqref{Kasympsum} equivalently as
\be\label{Kasympsum2}
K(t,x,y)\sim \frac{e^
{-m^2 t}}{4\pi t} \Big[ e^{-\ell^2(x,y)/4t} \sum_{k\ge 0} t^k a_k(x,y)+  \sum_{\del\cM_i}
 e^{-\ell^2_{i}(x,y)/4t} \sum_{k\ge 0} t^k  a_k^{(i)}(x,y) \Big]  \ .
 \ee

%Then, for small $t$, the heat kernel $K(t,x,y)$ is exponentially vanishing unless $x$ is close to $y$ or $x$ is close to any one $y_C^{(i)}$. The latter case is only possible if $x$ and $y$ are both close to the boundary component $\del\cM_i$. If $\ell(x,\del\cM)$ denotes the geodesic distance of $x$ to the boundary, we can phrase this by saying that all terms of the sum in the second term in \eqref{Kasympsum} are exponentially small if $\ell^2(x,\del\cM), \ell^2(y,\del\cM)\gg t$. Thus
%\be\label{Kasympbis}
%K(t,x,y) \sim \frac{e^{-m^2 t}}{4\pi t} e^{ -\ell^2(x,y)/4t } \sum_{k\ge 0} t^k a_k(x,y)\ ,
%\quad {\rm if}\ \ell^2(x,\del\cM)\gg t,\  \ell^2(y,\del\cM)\gg t\ ,
%\ee
%and in particular
%\be\label{Kasympxx}
%K(t,x,x) \sim \frac{e^{-m^2 t}}{4\pi t} \Big[ 1 + \sum_{k\ge 1} t^k a_k(x,x) \Big]\ ,
%\quad {\rm if}\ \ell^2(x,\del\cM)\gg t \ ,\ee
%with the same local expressions $a_r(x,x)$ as on a compact Riemann surface, e.g. $a_1(x,x)=\frac{R(x)}{6}$.
%However, as $x$ and $y$ get close to any given boundary component $\del\cM_i$, the corresponding ``image" term contributes and

For $x=y$ we have, in particular, $\ell^2_1(x,x)=4 \ell^2(x,\del\cM_i)$, where $\ell(x,\del\cM_i)$ denotes the geodesic distance of the point $x$ to the boundary $\del\cM_i$. Thus
\be\label{Kasympsum3}
K(t,x,x)\sim \frac{e^{-m^2 t}}{4\pi t} \Big[ \big(1 + \sum_{k\ge 1} t^k a_k(x,x) \big)+  \sum_{\del\cM_i}
e^{-\ell^2(x,\del\cM_i)/t} \big( 1 + \sum_{k\ge 1} t^k a_k^{(i)}(x,x) \big) \Big]   \ .
\ee
Here, the local expressions $a_r(x,x)$ are the same as on a compact Riemann surface without boundary, e.g. $a_1(x,x)=\frac{R(x)}{6}$.

If we are going to take the $t\to 0$ limit, we will find that the terms involving the boundaries drop out, unless the point $x$ is on the boundary $\del\cM_i$. In this case the corresponding boundary terms diverge for $t\to 0$ (as do the bulk terms). Thus these boundary terms behave as a Dirac delta concentrated on the boundary. To be more precise, let us look at  the heat kernel evaluated at $x=y$ and integrated over the manifold against a ``test function" $f$ ~:
\be\label{Kyyf}
\int_{\cM} \d^2 x \,\sqrt{g}\, f(x) K(t,x,x) \ .
\ee
Then the first term in \eqref{Kasympsum3} just gives the usual bulk result, while each of the boundary terms yields
\be\label{Imint1}
\frac{e^{-m^2 t}}{4\pi t}
\int_{\cM} \d^2 x \,\sqrt{g}\, f(x)  e^{-\ell^2(x,\del\cM)/t}  \big( 1 + \sum_{k\ge 1} t^k a_k^{(i)}(x,x) \big)  \ .
\ee
Again, for small $t$, the exponential forces $x$ to be close to the boundary. We may then view the integral as an integral over the boundary and an integral normal to the boundary. For a given boundary point $x_B$ we can Taylor expand all quantities around this point and do the integral in the normal direction. The leading small-$t$ term of this normal integral then simply is given by (using Riemann normal coordinates around $x_B$)
\be\label{normalint}
e^{-m^2 t}\int_0^\infty \d x_n \sqrt{\wh g(x_B)} f(x_B) \frac{e^{-(x-x_B)^2/t}}{4\pi t}=\frac{e^{-m^2 t}}{8\pi}  \sqrt{\frac{\pi}{t}}\sqrt{\wh g(x_B)} f(x_B) \ ,
\ee
where $\wh g$ is the metric induced on the boundary, so that $\d x_B \sqrt{\wh g(x_B)}=\d l$.
The  ${\cal O}(t^0)$-corrections to this expression involve the  normal derivatives of $\sqrt{g}$ and of $f$. As a result the small-$t$  asymptotic expansion of \eqref{Kyyf}  has the form
\be\label{Kyyf2}
\int_{\cM} \d^2 x \,\sqrt{g}\, f(x) K(t,x,x) 
=\frac{1}{4\pi t}\Bigg[ \int \d^2 x\, \sqrt{g(x)} f(x) + \frac{\sqrt{\pi t} }{2}\ \sum_i \int_{\del\cM_i} \d l\, f(x) +{\cal O}(t)\Bigg]
\ .
\ee
The leading small-$t$ singularity $\sim t^{-1}$ is given by the usual bulk term, while the boundary-terms yield subleading singularities $\sim t^{-1/2}$.

%\newpage

%%%%%%%%%%%%%%%%%%%%%%%%%%%%%%%%%%%%%%%%%%%%
\section{Local $\zeta$-functions and Green's function at coinciding points}\label{zetasec}

Recall that  local versions of the  $\zeta$-functions  were defined in \eqref{localzetadef} as
$\zeta(s,x,y)=\sum_{n\ge0} \frac{\vf_n(x) \vf_n(y) }{\l_n^s}$.
Note that $\zeta(1,x,y)=G(x,y)$.  They are related to the heat kernel by
\be\label{zetaheat}
\zeta(s,x,y)=\frac{1}{\G(s)} \int_0^\infty \d t\, t^{s-1} K(t,x,y) \ .
\ee
Of course, this formula involves the heat kernel for all values of $t$, not just the small-$t$ asymptotics. However,
for $s=0,-1,-2,\ldots$, $\frac{1}{\G(s)}$ has zeros and the value of $\zeta(s,x,y)$ is entirely determined by the singularities of the integral over $t$ that arise from the small-$t$ asymptotics of $K$. As shown above, the latter is given by local quantities on the Riemann surface. In particular, for any point not on the boundary of $\cM$, we have
\be\label{zetaat0}
\zeta(0,x,x)=\frac{R(x)}{24\pi}-\frac{m^2}{4\pi}  \quad , \quad x\ne\del\cM \ .
\ee
On the other hand, the values for $s=1,2,3,\ldots$ or the derivative at $s=0$ cannot be determined just from the small-$t$ asymptotics and require the knowledge of the full spectrum of $\D_g+m^2$, i.e. they contain global information about the Riemann surface.

Clearly, $\zeta(1,x,y)=G(x,y)$ is singular as $x\to y$. For $s\ne 1$,  $\zeta(s,x,y)$  provides a regularization of the propagator. It will be useful to study in more detail the singularities of $\zeta(s,x,y)$ which occur for $s\to 1$ \underline{and} $x\to y$. More generally, as is clear from \eqref{zetaheat}, any possible singularities of $\zeta(s,x,y)$ for $s\le 1$ come from the 
region of the integral where $t$ is small. Thus, we  tentatively let
\be\label{zetadecomp}
%\zeta_{\rm R}(s,x,y)=\zeta(s,x,y)-\zeta_{\rm sing}(s,x,y) \ , \quad 
\zeta_{\rm sing}(s,x,y)\simeq\frac{1}{\G(s)} \int_0^{\m^{-2}} \d t\, t^{s-1} K(t,x,y) \ ,
\ee
where $\m$ is some (arbitrary) large scale we introduce to separate the singular and non-singular parts, so that $\zeta-\zeta_{\rm sing}$ is free of singularities. For large $\m^2$, say $\m^2 A\gg 1$, where $A$ is the area of our manifold, we can use the small-$t$ asymptotics \eqref{Kasympsum} or \eqref{Kasympsum2} of $K$ to evaluate $\zeta_{\rm sing}$. With $t$ small, the $e^{-\ell^2/4t}$ are exponentially small unless $\ell^2 \lesssim t$. This means that in the first sum we must have $y=x+{\cal O}(\sqrt{t})$ and in the second sum $y_C^{(i)}=x+{\cal O}(\sqrt{t})$. Since $a_0(x,y)=a_0(x,x)+{\cal O}(\ell^2(x,y)\, R)=1+{\cal O}(t\, R)$, and similarly for $a_0^i(x,y_C^{(i)})$, and since the ${\cal O}(t R)$ terms do not contribute to the singularity at $s\to 1$, we define $\zeta_{\rm sing}(s,x,y)$ more precisely as 
\ba\label{zetasing}
\zeta_{\rm sing}(s,x,y)&=&\frac{1}{4\pi\, \G(s)} \int_0^{\m^{-2}} \d t\, t^{s-2} \big( e^{-\ell^2(x,y)/4t}
+\sum_{\del\cM_i}e^{-\ell^2_i(x,y)/4t}\big)  \nonumber\\
&=& \frac{\m^{2-2s}}{4\pi \G(s)} \Big[  E_s\big(\frac{\ell^2(x,y)\m^2}{4}\big)
+\sum_{\del\cM_i}\ E_s\big(\frac{\ell^2_i(x,y)\m^2}{4}\big) \Big]
\ ,
\ea
where the exponential integral (or incomplete gamma) function is defined by
\be\label{Enfct}
E_r(z)=\int_1^\infty \d u\, u^{-r} e^{-z u} \ , \quad \frac{\d}{\d z} E_r(z) =- E_{r-1}(z)\ .
\ee
As $z\to 0$, the $E_r(z)$ are regular for $r>1$ and have a logarithmic singularity for $r=1$ (see the appendix):
\be\label{E1sing}
E_1(z)=-\g -\log z +{\cal O}(z) \quad {\rm as}\  z\to 0 \  .
\ee
For $x\ne y$, the exponential integral functions are non-singular and we can set $s=1$ in \eqref{zetasing},
\be\label{zetasings1}
\zeta_{\rm sing}(1,x,y)= \frac{1}{4\pi} \Big[  E_1\big(\frac{\ell^2(x,y)\m^2}{4}\big)
+\sum_{\del\cM_i}\ E_1\big(\frac{\ell^2_i(x,y)\m^2}{4}\big) \Big]
\ ,
\ee
with the singularity appearing as the short-distance singularity for $x\to y$. We have :
\be\label{zetasings1gen}
{\rm as}\ x\to y \ : \quad
\zeta_{\rm sing}(1,x,y)\simeq \frac{1}{4\pi} \Big[  -\g-\log\big(\frac{\ell^2(x,y)\m^2}{4}\big)
+\sum_{\del\cM_i}\ E_1\big(\frac{\ell^2_i(x,y)\m^2}{4}\big)  \Big] \ ,
\ee
up to terms that vanish for $x=y$. If moreover $x\to y \to \del\cM_i$, i.e. they go to one of the boundaries, one has (again, up to terms that vanish in this limit)
\be\label{zetasings1bound}
{\rm as}\ x\to y\to \del\cM_i\ : \quad
\zeta_{\rm sing}(1,x,y)\simeq \frac{1}{4\pi} \Big[  -2\g-\log\big(\frac{\ell^2(x,y)\m^2}{4}\big)
-\log\big(\frac{\ell^2_i(x,y)\m^2}{4}\big)  \Big] \ .
\ee

On the other hand, for $s\ne 1$, we can set $x=y$ directly in \eqref{zetasing}. 
More precisely, we assume $\Re s>1$ and analytically continue in the end.
Then (recall $\ell^2_i(y,y)=4\ell^2(y,\del\cM_i$))
\be\label{zetasingx=y}
\zeta_{\rm sing}(s,y,y)= \frac{\m^{2-2s}}{4\pi \G(s)} \Big[  \frac{1}{s-1} 
+\sum_{\del\cM_i}\ E_s\big(\ell^2(y,\del\cM_i)\m^2\big) \Big] \ .
\ee
If $y\notin \del\cM$, only the first term yields a pole at $s=1$, while for $y\in\del\cM$ the second term also yields the same pole and, hence, the residue is doubled:
\be\label{zetasingx=y=yC}
\zeta_{\rm sing}(s,y,y)= \frac{\m^{2-2s}}{4\pi \G(s)} \  \frac{2}{s-1} 
 \ , \quad (y\in\del\cM) \ .
\ee
Thus, we see that  $\zeta(s,x,x)$ has a pole at $s=1$ with residue $\frac{a_0(x,x)}{4\pi}=\frac{1}{4\pi}$ for $x\notin\del\cM$ and residue $\frac{a_0(x,x)}{2\pi}=\frac{1}{2\pi}$ for $x\in\del\cM_i$.

Just as for the heat kernel itself, we will actually encounter expressions where $\zeta_{\rm sing}(s,y,y)$ is multiplied by some $f(y)$ and integrated over the manifold. Proceeding similarly to the derivation of 
\eqref{Kyyf2} we find
\ba\label{zetaf}
\hskip-1.cm\int_{\cM} \d^2 y \sqrt{g} f(y) \zeta_{\rm sing} (s,y,y) &=& \frac{\m^{2-2s}}{4\pi\G(s)} \frac{1}{s-1} \int_{\cM} \d^2 y \sqrt{g} f(y) + \frac{\m^{1-2s}}{8\sqrt{\pi}  \G(s)} \frac{1}{s-\frac{1}{2}} \int_{\del\cM} \hskip-2.mm \d l\, f(y) \nonumber\\
&&\hskip-1.cm+ \frac{\m^{-2s}}{8\pi  \G(s)} \frac{1}{s} \int_{\del\cM}\hskip-2.mm \d l\, \del_n f(y) 
+ \frac{\m^{-1-2s}}{32\sqrt{\pi} \G(s)} \frac{1}{s+\frac{1}{2}} \int_{\del\cM}\hskip-2.mm \d l\, \del_n^2 f(y)  +\ldots
\ea
This integrated expression exhibits poles at $s=1, \frac{1}{2}, -\frac{1}{2}, -1, -\frac{3}{2},\ldots$ and {\it no} pole at $s=0$. This infinite series of poles translates the discontinuous behaviour  between \eqref{zetasingx=y} and \eqref{zetasingx=y=yC} due to the fact that the limits $s\to 1$ and $z\to 0$ of $E_s(z)$ do not commute, as detailed in the appendix.
In particular, one has
\ba\label{zetazetaprime}
&&\lim_{s\to 1}  \Big[ 1+(s-1) \big(\frac{\d}{\d s} +\log\wh\m^2\big)\Big]
\int_{\cM}\hskip-2.mm \d^2 y \sqrt{g}\, f(y)\,\zeta_{\rm sing} (s,y,y) 
\nonumber\\
&&\hskip3.cm =\ \frac{1}{4\pi} \big(\g-\log\frac{\m^2}{\wh\m^2} \big) \int_{\cM}\hskip-2.mm \d^2 y \sqrt{g}\, f(y)
+\frac{1}{4\sqrt{\pi} \m}\int_{\del\cM}\hskip-3.mm \d l\, F(y,\m) \ ,
\ea
where
\be\label{Ffct}
F(y,\m)=f(y) + \frac{1}{2\sqrt{\pi}\m} \del_n f(y) + \frac{1}{12\m^2} \del_n^2 f(y) + \ldots \ .
\ee
All boundary terms are at least $\sim\frac{1}{\m}$ and we can thus restate the previous relation as
\be\label{zetazetaprimebis}
%&&
\lim_{s\to 1}  \Big[ 1+(s-1) \big(\frac{\d}{\d s} +\log\wh\m^2\big)\Big]
\int_{\cM} \d^2 y \sqrt{g}\, f(y)\,\zeta_{\rm sing} (s,y,y) 
%\nonumber\\
%&&\hskip2.cm 
=\frac{1}{4\pi}\big( \g-\log\frac{\m^2}{\wh\m^2}\big) \int_{\cM} \d^2 y \sqrt{g}\, f(y) +{\cal O}\big(\frac{1}{\m}\big) 
\\
\ee

In any case, 
\be\label{zetaregdef}
\zeta_{\rm R}(s,x,y)=\zeta(s,x,y)-\zeta_{\rm sing}(s,x,y)
\ee 
is free of singularities and, in particular,  has finite limits as $s\to 1$ and $x\to y$, in one order or the other,  i.e. $\zeta_{\rm R}(1,x,x)$ is finite and well-defined. We then let
\be\label{Gzetazetadef}
G_\zeta(y)=\zeta_{\rm R}(1,y,y)+\frac{\g}{4\pi}
=\lim_{s\to 1} \ \big(\zeta(s,y,y)-\zeta_{\rm sing}(s,y,y) \big)+\frac{\g}{4\pi}
\ee
This is an  important quantity, called the ``Green's function at coinciding points". Note that $G_\zeta(y)$   contains global information about the Riemann surface and cannot be expressed in terms of local quantities only.
Combining \eqref{zetazetaprime} and \eqref{Gzetazetadef} we get
\ba\label{zetazetaprimefull}
&&\lim_{s\to 1}  \Big[ 1+(s-1)  \big(\frac{\d}{\d s} +\log\wh\m^2\big) \Big]
\int_{\cM}\hskip-2.mm \d^2 y \sqrt{g}\, f(y)\,\zeta(s,y,y) 
\nonumber\\
&&\hskip2.cm =  \int_{\cM}\hskip-2.mm \d^2 y \sqrt{g}\, f(y)\, \big(G_\zeta(y)  - \frac{1}{4\pi}\log\frac{\m^2}{\wh\m^2} \big)
 +\frac{1}{4\sqrt{\pi} \m}\int_{\del\cM}\hskip-3.mm \d l\, F(y,\m) \ .
\ea
Note that the precise definition of $G_\zeta$ depends on our choice of $\m$, as is also obvious from this last relation since its left-hand side is $\m$-independent.

Maybe it is useful to pause and comment on the role of $\m$. It was introduced to separate $\zeta$ into its singular and regular parts. One might  thus view it as some sort of UV-cutoff. But contrary to a usual UV-cutoff, our formula are valid for any finite $\m$ and the relevant quantities that will appear in the gravitational action, such as \eqref{zetazetaprimefull} do not depend on the value of $\m$. One might then take the limit $\m\to\infty$ to simplify the formula, but one must be aware that $G_\zeta$ itself does not have a well-defined limit, only the combination $G_\zeta(y)  - \frac{1}{4\pi}\log\frac{\m^2}{\wh\m^2}$ does.

The other ingredient needed for computing the variation of the gravitational action was\break\hfill
$ \lim_{s\to 0} \Big[ 1+s\big(\frac{\d}{\d s} +  \log\wh\m^2\big)\Big] \int\d^2 x\sqrt{g}\,  \dd\s(x) \zeta(s,x,x)$.
Again, one sees from \eqref{zetaf}, by replacing $\zeta_{\rm sing}(s,x,x)$ by $\zeta(s,x,x)$ that $\int\d^2 x\sqrt{g}\,  \dd\s(x) \zeta(s,x,x)$ actually has poles for $s=1, \frac{1}{2}, -\frac{1}{2}, -1, \ldots$ but {\it not} for $s=0$, since the would-be pole is cancelled by the $\frac{1}{\G(s)}$. Hence, $\int\d^2 x\sqrt{g}\,  \dd\s(x) \zeta(s,x,x)$ is regular at $s=0$ and, adding the bulk contribution \eqref{zetaat0} and the boundary contribution read from \eqref{zetaf}, we get
\ba\label{zetazetaprimes0}
\lim_{s\to 0} \Big[ 1+s\big(\frac{\d}{\d s} +  \log\wh\m^2\big)\Big] \int\d^2 x\sqrt{g}\,  f(x) \zeta(s,x,x)
&=&\int\d^2 x\sqrt{g}\,  f(x) \zeta(0,x,x)
\nonumber\\
&&\hskip-5.cm =\ \frac{1}{4\pi}\int\d^2 x\sqrt{g}\,  f(x) \big( \frac{R}{6} -m^2\big) +\frac{1}{8\pi}\int_{\del\cM} \d l\, \del_n f(x) \ .
\ea

Let us relate $G_\zeta(y)$ to the  Green's function $G(x,y)$ at coinciding points with the short-distance singularity subtracted.
Since $\zeta_{\rm R}(s,x,y)=\zeta(s,x,y)-\zeta_{\rm sing}(s,x,y)$ is free of singularities, we may change the order of limits. If we first let $s=1$, so that $\zeta(1,x,y)=G(x,y)$ and $\zeta_{\rm sing}(1,x,y)$ is given by \eqref{zetasings1gen}, we find
\be\label{Gzeta1}
G_\zeta(y)
=\lim_{x\to y} \left[ G(x,y) + \frac{1}{4\pi}\Big( \log\big(\frac{\ell^2(x,y)\m^2}{4}\big) + 2\g
-\sum_{\del\cM_i}\ E_1\big(\frac{\ell^2_i(x,y)\m^2}{4}\big) \Big) \right] \ .
\ee
We know that $G_\zeta(y)$ is a non-singular quantity for all $y\in \cM$, in particular also on the boundary. The logarithm subtracts the generic short-distance singularity of $G(x,y)$, while the $E_1$ subtract the additional singularities present whenever $y\in \del\cM_i$.
%If we take furthermore $y$ to the boundary component $\del\cM_i$ then
%\be\label{Gzeta2}
%G_\zeta(y)=\lim_{x\to y} \left[ G(x,y) + \frac{1}{4\pi}\left( \log\big(\frac{\ell^2(x,y)\m^2}{4}\big) 
%+\log\big(\frac{\ell^2(x,y_C^{(i)})\m^2}{4}\big) + 3\g\right) \right] \ , \ {\rm as} \ y\to y_C \ .
%\ee
%The same relations hold when the zero-mode is excluded, i.e.~between $\wt G_\zeta(y)$, $\wt G(x,y)$ and $\wt \zeta(s,y,y)$.  
If, as before, we multiply this relation by some smooth $f(y)$ and integrate over the manifold, we get in particular for these $E_1$-terms:
\ba
\lim_{x\to y} \int_{\cM}\hskip-2.mm \d^2 y \sqrt{g}\,f(y) E_1\big(\frac{\ell^2_i(x,y)\m^2}{4}\big) 
\hskip-2.mm&=&\hskip-2.mm \int_0^{\m^{-2}} \frac{\d t}{t} \int_{\cM}\hskip-2.mm\d^2 y \sqrt{g}\, f(y)e^{-\ell^2(y,\del\cM_i)/t}
=\frac{\sqrt{\pi}}{\m} \int_{\del\cM_i}\hskip-3.mm\d l\, F(y,\m)  ,
\nonumber\\
&&
\ea
with $F(y,\m)$ defined in \eqref{Ffct}.
It follows that we may rewrite \eqref{zetazetaprimefull} as
\be\label{zetazetaprimefullbis}
\lim_{s\to 1}  \Big[ 1+(s-1) \big(\frac{\d}{\d s}+\log\wh\m^2\big) \Big]
\int_{\cM}\hskip-2.mm \d^2 y \sqrt{g}\, f(y)\,\zeta(s,y,y) 
=\int_{\cM}\hskip-2.mm \d^2 y \sqrt{g}\, f(y)\, G_{\rm R,bulk}(y) \ ,
\ee
where
\be\label{GRbulk}
G_{\rm R,bulk}(y)=
\lim_{x\to y} \Bigg[ G(x,y)   + \frac{1}{4\pi}\Big( \log\big(\frac{\ell^2(x,y)\wh\m^2}{4}\big) + 2\g\Big) \Bigg]
\ee
is the Green's function at coinciding points with its bulk singularity subtracted.
If necessary, \eqref{zetazetaprimefullbis} again shows that this does not depend on the arbitrarily introduced $\m$ (although it does depend on $\wh\m$ which was part of our definition of the functional integral). While the quantity $G_{\rm R,bulk}(x)$ has the advantage of being $\m$-independent, it has a (logarithmic) singularity as $x$ approaches the boundary. However, we know that these singularities must be integrable as is clear from the equality of \eqref{zetazetaprimefullbis} with \eqref{zetazetaprimefull} which is finite, independently of the arbitrary choice of $\m$. 
As will become clear next, while $G_\zeta(x)$ satisfies Neumann boundary conditions, this is not the case of $G_{\rm R,bulk}(x)$.

To study the boundary condition satisfied by $G_\zeta(y)$  we only need its behaviour in the immediate vicinity of the relevant boundary component which can be read from  \eqref{Gzeta1}:
\be\label{Gzeta2}
G_\zeta(y)
\simeq\lim_{x\to y} \left[ G(x,y) + \frac{1}{4\pi}\Big( \log\big(\frac{\ell^2(x,y)\m^2}{4}\big) 
+ \log\big(\frac{\ell^2_i(x,y)\m^2}{4}\big) + 3\g \Big) \right]  \ , \quad {\rm as}\ y\to \del\cM_i \ .
\ee
Now, $G(x,y)$ satisfies the Neumann condition in both its arguments. The same is true for the sum of the logarithms,  up to terms that vanish as $x$ and $y$ approach the boundary. This can be seen as follows:
As one zooms in close to the boundary, the boundary becomes flat and the geometry locally Euclidean, and using Riemann normal coordinates in the normal and tangential directions around the relevant boundary point (such that the boundary is at zero normal coordinate), one has $\ell^2(x,y)\simeq (x_t-y_t)^2+(x_n-y_n)^2$ as well as $\ell^2_i(x,y)\simeq (x_t-y_t)^2+(x_n+y_n)^2$. Then $\del_{x^n}\ell^2(x,y)\vert_{x^n=0}\simeq -2y_n=-\del_{x^n}\ell^2_i(x,y)\vert_{x^n=0}$, as well as $\ell^2(x,y)\vert_{x^n=0}\simeq \ell^2_i(x,y)\vert_{x^n=0}$. It follows that $\del_{x^n} \big[\log\big(\frac{\ell^2(x,y)m^2}{4}\big) +\log\big(\frac{\ell^2_i(x,y)\m^2}{4}\big)\big]\vert_{x^n=0}=0$, i.e. the sum of the logarithms  satisfies the Neumann condition in $x$ up to terms that vanish as $x$ and $y$ approach the boundary. Since $\ell^2_i(x,y)$ is symmetric in $x$ and $y$, the same is true in $y$.
Now if any function $h(x,y)$ satisfies the Neumann condition in both arguments, the function $H(y)=\lim_{x\to y} h(x,y)=\lim_{\e\to 0} h(y+\e,y)$ then obviously also satisfies the Neumann condition. We conclude that $G_\zeta$ satisfies the Neumann boundary condition on every boundary component $\del\cM_i$, i.e.
\be\label{GzetaNeumann}
n^a \del_a G_\zeta(y) =0 \ , \quad {\rm for}\ y\in\del\cM \ .
\ee
It follows that, if $\f$ is any smooth function that also satisfies Neumann conditions, one has
\be\label{partintGzeta}
\int_{\cM} \d^2 y \sqrt{g} \, \D \f\, G_\zeta(y) = \int_{\cM} \d^2 y \sqrt{g}\,  \f\,  \D G_\zeta(y) \ .
\ee
It is now also clear that $G_{\rm R,bulk}$ does not satisfy the Neumann condition since its definition lacks the crucial third term in \eqref{Gzeta2}.

We can now evaluate \eqref{zetazetaprimefull} for $f=\D \f$ and use \eqref{partintGzeta} to get
\ba\label{phideltaGzeta}
&&\lim_{s\to 1}  \Big[ 1+(s-1)\big( \frac{\d}{\d s} +\log\wh\m^2\big) \Big]
\int_{\cM}\hskip-2.mm \d^2 y \sqrt{g}\, \D \f(y)\,\zeta(s,y,y) 
\nonumber\\
&&\hskip 4.cm=  \int_{\cM}\hskip-2.mm \d^2 y \sqrt{g}\, \f(y)\, \D G_\zeta(y) +\frac{1}{4\sqrt{\pi} \m}\int_{\del\cM}\hskip-3.mm \d l\, \Phi(y,\m)  \ , ,
\ea
where $\Phi(y,\m) = \D \f +\frac{1}{2\sqrt{\pi}\m} \del_n \D\f +\frac{1}{12\m^2} \del_n^2\D\f+\ldots$.
At this point one might be tempted to take $\m\to \infty$ to get rid of the last term but, of course, one must remember that   $G_\zeta$ also depends on $\m$. However, this relation shows that, since the left-hand side does not depend on $\m$, the quantity $\int_{\cM} \d^2 y \sqrt{g}\, \f(y)\, \D G_\zeta(y)$ has a finite limit as $\m\to \infty$ and we arrive at the two following equivalent expressions:
\ba\label{phideltaGzetabis}
&&\hskip-3.cm \lim_{s\to 1}  \Big[ 1+(s-1) \big( \frac{\d}{\d s} +\log\wh\m^2\big)  \Big]
\int_{\cM}\hskip-2.mm \d^2 y \sqrt{g}\, \D \f(y)\,\zeta(s,y,y) \nonumber\\
&&\hskip3.cm = \lim_{\m\to \infty} \int_{\cM}\hskip-2.mm \d^2 y \sqrt{g}\, \f(y)\,  \D G_\zeta(y) 
\nonumber\\
&&\hskip3.cm = \int_{\cM}\hskip-2.mm \d^2 y \sqrt{g}\, \D\f(y)\,  G_{\rm R,bulk}(y)\big) \ .
\ea
Both ways of writing require a comment:
While the $\m\to 0$ limit of the integral involving $\D G_\zeta$ exists,  this is not the case of $\D G_\zeta(y)$ itself for $y$ on the boundary.
On the other hand, in the integral invoving $G_{\rm R,bulk}$, even though $G_{\rm R,bulk}$ does not satisfy the Neumann condition, one might want to integrate by parts generating a boundary term:
\be\label{wrongequation}
\int_{\cM}\hskip-2.mm \d^2 y \sqrt{g}\, \D\f(y)\,  G_{\rm R,bulk}(y)\big)  \ ? \ = \ ? \
\int_{\cM}\hskip-2.mm \d^2 y \sqrt{g}\, \f(y)\,  \D G_{\rm R,bulk}(y)\big) + \int_{\del\cM} \d l\, \f(y) \del_n G_{\rm R,bulk}(y) \ .
\ee
However, this is not possible : both terms on the r.h.s. are meaningless since $\D G_{\rm R,bulk}(y)$ has a non-integrable singularity as $y$ approaches the boundary (expected to be $\sim 1/\ell^2(y,\del\cM)$), and $\del_n G_{\rm R,bulk}$ is infinite everywhere on the boundary.

%%%%%%%%%%%%%%%%%%%%%%%%%%%%%%%%%%%%%%%%%%%%

\section{The Mabuchi action on a manifold with boundaries}

We are now in position to assemble our results and determine the gravitational action on a Riemann surface with boundaries.
As already explained, the strategy is to use the infinitesimal variation of $S_{\rm grav}$ under an infinitesimal change of the metric as given by \eqref{deltaSgrav}, and then to integrate $\dd S_{\rm grav}$ to obtain $S_{\rm grav}[g,g_0]$.

Inserting \eqref{zetazetaprimes0} and \eqref{zetazetaprimefullbis} into \eqref{deltaSgrav}, we immediately get
\ba\label{deltaSgrav2}
\dd S_{\rm grav}&=& -\frac{1}{24\pi}\Big[ \int_{\cM} \sqrt{g}  \, \dd\s\, R + 3 \int_{\del\cM}\hskip-2.mm  \d l\, \del_n \dd\s \Big]
+\, m^2  \Big[  \int_{\cM} \sqrt{g}  \, \dd\s\,  \big( G_\zeta+\frac{1}{4\pi}\big) +\frac{1}{4\sqrt{\pi}\m} \int_{\del\cM}\hskip-2.mm \d l \ \Sigma(\m) \Big] 
\nonumber\\
&=& -\frac{1}{24\pi}\Big[ \int_{\cM} \sqrt{g}  \, \dd\s\, R + 3 \int_{\del\cM}\hskip-2.mm  \d l\, \del_n \dd\s \Big]
+\, m^2   \int_{\cM} \sqrt{g}  \, \dd\s\,  \big( G_{\rm R,bulk}+\frac{1}{4\pi}\big)\  .
\ea
Note that this is not an expansion in powers of $m^2$ but an exact result. Our perturbation theory was a first order perturbation in $\dd\s$, not in $m^2$. Indeed, $G_\zeta$ and $G_{\rm R,bulk}$ still depend on $m^2$
and we get exactly the first two terms in an expansion in powers of $m^2$ if we replace them by the corresponding quantities $G_\zeta^{(0)}$ and  $G^{(0)}_{\rm R,bulk}$ defined for the massless case. However there is a subtlety here, since in the massless case the zero-mode must be excluded from the sum over eigenvalues defining the Green's function. If we denote with a tilde all quantities lacking the zero-mode contribution we have
\be\label{Greenzero}
G(x,y)=\frac{1}{m^2 A} +\wt G(x,y)
\quad , \quad
G_\zeta(x) =\frac{1}{m^2 A}+\wt G_\zeta(x)
\quad , \quad
G_{\rm R,bulk}(x) =\frac{1}{m^2 A}+\wt G_{\rm R,bulk}(x) \ .
\ee
The quantities $\wt G$, $\wt G_\zeta$  and $\wt G_{\rm R,bulk}$ all have a smooth limit as $m\to 0$. Thus, the expansion in powers of $m^2$ reads
\ba\label{deltaSgrav3}
\dd S_{\rm grav}&=& -\frac{1}{24\pi}\Big[ \int_{\cM} \sqrt{g}  \, \dd\s\, R + 3 \int_{\del\cM}\hskip-2.mm  \d l\, \del_n \dd\s 
\Big] +\frac{\dd A}{2A}
\nonumber\\
&&+\, m^2  \Big[  \int_{\cM} \sqrt{g}  \, \dd\s\,  \big( \wt G^{(0)}_\zeta+\frac{1}{4\pi}\big) +\frac{1}{4\sqrt{\pi}\m} \int_{\del\cM}\hskip-2.mm \d l \ \Sigma(\m) \Big]  \ +\, {\cal O}(m^4) 
\nonumber\\
&=& -\frac{1}{24\pi}\Big[ \int_{\cM} \sqrt{g}  \, \dd\s\, R + 3 \int_{\del\cM}\hskip-2.mm  \d l\, \del_n \dd\s 
\Big] +\frac{\dd A}{2A}
+\, m^2  \Big[  \int_{\cM} \sqrt{g}  \, \dd\s\,  \big( \wt G^{(0)}_{\rm R,bulk}+\frac{1}{4\pi}\big)  \ +\, {\cal O}(m^4)\ .
\nonumber\\
\ea
The first term is independent of $m$ and corresponds to $-\frac{1}{24\pi}$ times the variation of the Liouville action on a manifold with boundary\footnote{
With respect to the metrics $g_0$ and $g=e^{2\s}g_0$ one has $\d l=e^\s \d l_0$ as well as $n^a=e^{-\s}n^a_0$ and thus also $\del_n\equiv n^a\del_a=e^{-\s} \del_n^0$. One sees that $\d l\, \del_n \dd\s=e^\s \d l_0\, e^{-\s} \del_n^0 \dd\s=\d l_0\, \del_n^0\dd\s=\dd (\d l_0\, \del_n^0\s)=\dd(\d l\, \del_n\s)$.
}:
\be\label{SLiouville}
S_L=\int_{\cM} \sqrt{g_0}(g_0^{ab} \del_a \s \del_b \s+R_0 \s) + 3 \int_{\del\cM} \d l\, \del_n \s \ ,
\ee
while the $\frac{\dd A}{2A}$-term contributes a piece $\frac{1}{2}\log\frac{A}{A_0}$ to $S_{\rm grav}$. Recall that, contrary to $\f$ or $\dd\f$, the field $\s$ and its variation $\dd\s$ do not satisfy the Neumann condition.

To go further, we need the variation of $\wt G^{(0)}_\zeta$ and of $\int_{\del\cM} \d l \, \Sigma$  or the variation of  $\wt G^{(0)}_{\rm R,bulk}$ under an infinitesimal variation of the metric corresponding to $\dd\s$. 
At this point it turns out to be easier\footnote{
The relevant formulae for studying the variation of $\wt G^{(0)}_\zeta$ and of $\int_{\del\cM} \d l \, \Sigma$  are given in an appendix.
} 
to study the variation of $\wt G^{(0)}_{\rm R,bulk}$. The latter is obtained exactly as for a manifold without boundary. The simplest derivation just uses the differential equation satisfied by $G(x,y)$ to obtain
\be\label{Gvar}
\dd G(x,y) =-2 m^2\int \d^2 z \sqrt{g}\, G(x,z)\, \dd\s(z)\, G(z,y) \ ,
\ee
which  satisfies the Neumann conditions. Alternatively, one can use the perturbation theory formulae \eqref{deltalambda} and \eqref{deltaphi} to obtain
\ba\label{deltaG}
\dd G(x,y)&=& \sum_n \frac{\dd\vf_n(x) \vf_n(y) + \vf_n(x) \dd\vf_n(y)}{\l_n}-\frac{\vf_n(x)\vf_n(y) \dd\l_n}{\l_n^2}
\nonumber\\
&=& -2 m^2 \sum_{n,k} \frac{\langle \vf_k|\dd\s|\vf_n\rangle \vf_k(x) \vf_n(y)}{\l_n\l_k}
=-2 m^2\int \d^2 z \sqrt{g}\, G(x,z) \dd\s(z) G(z,y) \ ,
\ea
in agreement with \eqref{Gvar}. Next, the variation of $\ell^2(x,y)$ was given e.g.~in \cite{FKZ,BF,BLgrav}. In the limit $x\to y$ one simply has $\ell^2(x,y)\simeq g_{ab} \d x^a \d x^b= e^{2\s(y)} g^{(0)}_{ab} \d x^a \d x^b$ which shows that  one has
$\dd \ell^2(x,y) \simeq 2\dd\s(y)\, \ell^2(x,y)$ and, hence,
\be\label{ell2var}
\lim_{x\to y}\  \dd \log \big(\ell^2(x,y) \m^2\big) =2\, \dd\s(y) \ .
\ee 
It follows that
\be\label{GRvar}
\dd G_{\rm R,bulk}(x)=-2 m^2 \int \d^2 z \sqrt{g} \big( G(x,z)\big)^2\, \dd\s(z) + \frac{\dd\s(x)}{2\pi}  \ .
\ee
Separating the zero-mode parts $\frac{1}{m^2 A}$,  this is rewritten as
\be\label{GRvar2}
\dd \wt G_{\rm R,bulk}(x)=-\frac{4}{A} \int\d^2 z \sqrt{g} \,\wt G(x,z)\dd\s(z) -2 m^2 \int \d^2 z \sqrt{g} \big( \wt G(x,z)\big)^2\, \dd\s(z) 
+ \frac{\dd\s(x)}{2\pi}  \ .
\ee
Since $\dd\sqrt{g}=2\sqrt{g}\, \dd\s$, it follows that
\be\label{intGRvar}
\dd\int \sqrt{g} \, \wt G_{\rm R,bulk}(x)= \int \sqrt{g}\,  2\dd\s\, \big(\wt G_{\rm R,bulk}(x) +\frac{1}{4\pi} \big)
+ {\cal O}(m^2) \ .
\ee
(Note that the first term in \eqref{GRvar2} integrates to zero and does not contribute in \eqref{intGRvar}.)
Thus
\be\label{deltaSgrav4}
\dd S_{\rm grav}=-\frac{1}{24\pi} \dd S_{L} +\frac{1}{2}\frac{\dd A}{A}+ \frac{m^2}{2} \ \dd\   \int \sqrt{g}\, \wt G^{(0)}_{\rm R,bulk}(x) + {\cal O}(m^4)\ ,
\ee
where $\wt G^{(0)}_{\rm R,bulk}$ is computed from the Green's function without zero-mode of the massless theory.
The order $m^2$ term in \eqref{deltaSgrav4} is given by the variation of the functional
\be\label{Phi}
\Phi_G[g]=\int \sqrt{g}\, \wt G^{(0)}_{\rm R,bulk}(x;g) \ ,
\ee
where we explicitly indicated the dependence of $\wt G^{(0)}_{\rm R,bulk}$ on the metric $g$. Thus
\ba\label{Sgrav4}
 S_{\rm grav}[g,g_0]=-\frac{1}{24\pi}  S_{L}[g,g_0] +\frac{1}{2}\log\frac{A}{A_0}
+ \frac{m^2}{2} \, \Big(\Phi_G[g]-\Phi_G[g_0] \Big) + {\cal O}(m^4)\ .
\ea

In order to express $\Phi_G[g]-\Phi_G[g_0]$ as a local functional of $\s$ and $\f$, we use again  \eqref{GRvar2} in the zero-mass limit and replace $\dd\s$ in the first term by $\frac{\dd A}{2A}-\frac{A}{4}\D\dd\f$ according to \eqref{delsigmadelphirel}:
\ba\label{GRvar6}
\dd \wt G^{(0)}_{\rm R,bulk}(x)
&=& \int \d^2 z \sqrt{g} \,\wt G^{(0)}(x,z) \D \dd \f(z)  + \frac{\dd\s(x)}{2\pi}
\nonumber\\
&=&\dd\f(x)-\frac{1}{A} \int \d^2 z \sqrt{g} \, \dd\f(z)  + \frac{\dd\s(x)}{2\pi} 
 \ .
\ea
Note that we integrated the Laplace operator by parts without generating boundary terms since both $\wt G$ and $\dd\f$ satisfy the Neumann boundary conditions (which is not the case for $\dd\s$) and then we used the differential equation \eqref{Greeneigenmassless}. Equation \eqref{GRvar6} can be integrated as
\be\label{GRvarfinite}
\wt G^{(0)}_{\rm R,bulk}(x,g)-\wt G^{(0)}_{\rm R,bulk}(x,g_0)=
\f(x) +\frac{\s(x)}{2\pi}-S_{AY}[g_0,g] \ ,
\ee
with
\be\label{SAY}
S_{AY}[g_0,g]=-\int\sqrt{g_0}\, \Big( \frac{1}{4} g_0^{ab}\del_a\f \del_b\f - \frac{\f}{A_0}\Big) \ .
\ee
It is now straightforward to obtain
\ba\label{finitePhi}
\Phi_G[g]-\Phi_G[g_0]&=&\int \sqrt{g_0} \Big( \frac{A}{A_0} -\frac{A}{2}\D_0\f\Big)
\Big(\wt G^{(0)}_{\rm R,bulk}[g_0]+\f+\frac{\s}{2\pi} -S_{AY}[g_0,g]\Big)
-\Phi_G[g_0] \nonumber\\
&&\hskip-2.cm =\frac{A-A_0}{A_0} \Phi_G[g_0]
-\frac{A}{2}\int\sqrt{g_0}\Big(\frac{1}{2}\f\D_0\f  -\frac{1}{\pi A} \s e^{2\s} + \D_0\f\, \wt G^{(0)}_{\rm R,bulk}[g_0]  \Big)\ .
\ea
As already emphasized, contrary to $G(x,y)$ or $G_\zeta(x)$, the quantities $G_{\rm R,bulk}$ and $\wt  G^{(0)}_{\rm R,bulk}$ do not satisfy the Neumann condition. Moreover,  $\del_n G_{\rm R,bulk}$ and $\del \wt G^{(0)}_{\rm R,bulk}$ are singular on the boundary.

Thus, we arrive at
\ba\label{Sgrav5}
S_{\rm grav}[g,g_0]&=&-\frac{1}{24\pi}  S_{L}[g,g_0] +\frac{1}{2}\log\frac{A}{A_0}
+\frac{ m^2\, (A-A_0)}{2 A_0}   \Phi_G[g_0] 
\nonumber\\
&&+ \frac{m^2 A}{4} \, \Bigg[ \int\sqrt{g_0}\Big(-\frac{1}{2}\f\D_0\f  +\frac{1}{\pi A} \s e^{2\s} -\D_0\f\,  \wt G^{(0)}_{\rm R,bulk}[g_0] \Big)  \Bigg]
+ {\cal O}(m^4)\ .
\ea
(Recall our notation: a tilde on any quantity means that we removed the contribution of the zero-mode, and a superscript $(0)$ means that we are computing in the massless limit. So $\wt G^{(0)}_{\rm R, bulk}$ is $G_{\rm R,bulk}$ computed in the massless limit with the contribution of the zero-mode removed. The same remark applies to $\wt G^{(0)}_\zeta$ used below.)
The first line contains the usual Liouville action along with a factor $+\frac{1}{2}\log\frac{A}{A_0}$, as well as a contribution to the cosmological constant action.  The cosmological constant action is required in any case to act as a counterterm to cancel the divergence that accompanies the $\zeta'(0)$ when properly evaluating the determinant, e.g.~with the spectral cut-off regularization as was done in \cite{BF}. The terms in the second line are the genuine order $m^2$ corrections.   
Using the second equality in \eqref{phideltaGzetabis}, we can rewrite the latter using $\wt G^{(0)}_\zeta\equiv \wt G^{(0)}_\zeta[g_0,\m]$ instead of $\wt G^{(0)}_{\rm R,bulk}$. In particular, this allows us to integrate by parts the Laplacian and  to take the $\m\to\infty$ limit. Recall that $\m$ was arbitrary and our equations are valid for all values of $\m$. However, $G_\zeta$ does not have a well-defined $\m\to\infty$ limit, but $G_\zeta(y)  - \frac{1}{4\pi}\log\frac{\m^2}{\wh\m^2}$ does, and so does $\D_0 G_\zeta$, as well as $\wt G^{(0)}_\zeta$.  We then get
\ba\label{Sgrav7}
\hskip-1.cm S_{\rm grav}[g,g_0]&=&-\frac{1}{24\pi}  S_{L}[g,g_0] +\frac{1}{2}\log\frac{A}{A_0}
+\frac{ m^2\, (A-A_0)}{2 A_0}  \Phi_G[g_0]
\nonumber\\
&&+ \frac{m^2 A}{4} \,  \lim_{\m\to\infty} \Bigg[ \int\sqrt{g_0}\Big(-\frac{1}{2}\f\D_0\f  +\frac{1}{\pi A} \s e^{2\s} - \f\,  \D_0\wt G^{(0)}_\zeta [g_0,\m] \Big)  \Bigg]
+ {\cal O}(m^4)\ .
\ea
Written this way, the order $m^2$-terms ressemble the  usual Mabuchi plus Aubin-Yau actions found for manifolds without boundary \cite{FKZ,BLgrav}. However, here  the function $\D_0 \wt G^{(0)}_\zeta(x)$ no longer is a simple expression but depends non-trivially on the point $x$ and in particular on the distances from the various boundary components. Of course, the same is true for  $\wt G^{(0)}_{\rm R,bulk}(x)$.

%%%%%%%%%%%%%%%%%%%%%%%%%%%%%%%%%%%%%%%%%%%%

\section{The cylinder\label{seccyl}}

In this section we work out the gravitational action for the simplest two-dimensional manifold with a boundary : the cylinder. As we have seen in section \ref{cylexample} the heat kernel and hence also the Green's function on the cylinder are obtained from the corresponding quantities on the torus by a method of images. Thus to get the Green's function of the Laplace operator on the cylinder of length $T$ and circumference $2\pi R$ we first determine the Green's function on the torus with  periods  $2T$ and $2\pi R$, i.e.~modular parameter $\tau=i\pi\frac{R}{T}$. Actually, it is not more complicated to obtain the Green's function for a torus with arbitrary modular parameter $\tau$, but since we will be only interested in the ``straight" cylinder, we will explicitly consider the square torus with purely imaginary $\tau$.

With respect to our general notation, throughout this section we consider a fixed reference metric $g_0$ and corresponding Laplacian $\D_0$ and Green's functions $G(z_1,z_2;g_0)$ although we will mostly drop the reference to $g_0$.

\subsection{Green's function on the torus}

To get the Green's function on the torus with  periods  $2a$ and $2b$ , in principle, one could take the heat kernel ${\cal K}_{\rm torus}(t,x_1,x_2,y_1,y_2)$ as constructed from the eigenfunctions \eqref{cyleigenfct} and eigenvalues of the Laplace operator, cf \eqref{cylK} and integrate over $t$ form 0 to $\infty$. However, we have not been able to find any useful formula for $\int_0^\infty \d t\,  \t_3\big( i\frac{\pi t}{b^2}\big\vert \frac{x_2-y_2}{2b}\big) \t_3\big( i\frac{\pi t}{a^2}\big\vert \frac{x_1-y_1}{2a}\big)$. Instead, we will follow the usual approach to identify a suitable doubly periodic solution of the Laplace equation with the correct singularity at the origin. It will be convenient to use a complex coordinate $z$.  Thus, in this section we will change our notation with respect to the previous one and call $x$ and $y$ the real and imaginary parts of $z$:
\be
z=x+i y \quad , \quad x\simeq x+2a \quad , \quad y\simeq y+2b \ .
\ee
When we need to label two points\footnote{
This amounts to the substitutions $x_1\to x_1,\  x_2\to y_2, \ y_1\to x_2,\  y_2\to y_2$.
}, 
we will use $z_1=x_1+i y_1$ and $z_2=x_2+i y_2$.
We thus have a square torus with modular parameter $\tau=i\frac{b}{a}$.    The reference metric $g_0$ is just the standard metric $\d s^2 =\d z \d \zb$ and $\D_0=-4\del_z \del_{\zb}$.  

The Green's function ${\cal G}(z_1,z_2)$ must be doubly periodic in both $z_1$ and $z_2$ with periods $2a$ and $2b\, i$. By the translational invariance of the torus, it can only depend on the difference $z_1-z_2$, and it must exhibit the appropriate $-\frac{1}{4\pi}\log |z_1-z_2|^2$ singularity as $z_1\to z_2$ in order to satisfy $\D_0 {\cal G}(z_1,z_2)=\dd^{(2)}(z_1-z_2)-\frac{1}{A_0}$. This fixes ${\cal G}$ only up to some additive constant. The latter must be fixed such that $\int \d^2 z_1 {\cal G}(z_1,z_2)=\int \d^2 z_2 {\cal G}(z_1,z_2)=0$. Define the function $g$ (not to be confused with the metric) as (see e.g~\cite{KlargeN})
\be\label{torusg}
g(z)
%&=& \frac{(\Im z)^2}{8ab}-\frac{1}{4\pi} \log\Big[\t_1\big(\frac{z}{2a}\big\vert i\frac{b}{a}\big) \t_1\big(\frac{\zb}{2a}\big\vert i\frac{b}{a}\big) \Big]+\frac{1}{2\pi} \log \Big\vert\eta\big( i\frac{b}{a}\big)\Big\vert  \nonumber\\
=\frac{(\Im z)^2}{8ab}-\frac{1}{4\pi} \log\Big\vert \frac{\t_1\big(\frac{z}{2a}\big\vert i\frac{b}{a}\big)}{\eta\big( i\frac{b}{a}\big)}\Big\vert^2  \ ,
\ee
where the elliptic theta function $\t_1$ and the Dedekind $\eta$-function are defined as \cite{Erdelyi}
\ba\label{theta1def}
\t_1(\n\vert\tau)&=& 2 q^{1/4} \sum_{n=0}^\infty (-)^n q^{n(n+1)} \sin(2n+1)\pi \n
\nonumber\\
&=&2 q^{1/4} \sin\pi\n \prod_{n=1}^\infty (1-q^{2n})(1-2 q^{2n}\cos 2\pi\n + q^{4n})
\ , \quad q=e^{i\pi\tau}\ ,
\\
\eta(\tau)&=&q^{\frac{1}{12}} \prod_{n=1}^\infty (1-q^{2n}) \ .
\ea
$\t_1$ satisfies $\t_1(\n+1|\tau)=-\t_1(\n|\tau)$ and $\t_1(\n+\tau|\tau)=-e^{-i\pi(2\n+\tau)}\t_1(\n|\tau)$. It follows that
\begin{itemize}
\item
$g(x+iy)$ is periodic under $x\to x+2a$ and under $y\to y+2b$. 
\item
It is obvious from the factorization of the logarithm that 
\be\label{Laplg}
{\rm for}\ z\ne 0 \quad : \quad -4\del_z\del_{\zb} g=-\frac{1}{4ab}=-\frac{1}{A_0} \ ,
\ee 
where $A_0$ is the area of the torus. 
\item
As $z\to 0$ one has
\be\label{gzto0}
g(z)\sim -\frac{1}{4\pi}\log |z|^2-\frac{1}{2\pi} \log\Big( \frac{\pi}{a} q^{1/6} \prod_{n=1}^\infty (1-q^{2n})^2 \Big)
+{\cal O}(|z|) \ , \quad q=e^{-\pi b/a} \ .
\ee 
which together with the previous relation ensures that
\be\label{gisGreen}
\Delta_0 g(z) = - 4\del_z \del_{\zb} g(z)=\dd^{(2)}(z) -\frac{1}{A_0}\ .
\ee
\item
One can show that 
\be\label{gint0}
\int
d^2 z\, g(z) = 0 \ .
\ee
\item
We have the symmetry properties
\be\label{gsymprop}
g(\zb)=g(z) \quad , \quad g(-z)=g(z) \quad , \quad g'(-z)=-g'(z) \ .
\ee
\end{itemize}
Thus 
\be\label{Greentorus}
{\cal G}(z_1,z_2)=g(z_1-z_2)
\ee 
is the appropriate Greens's function on the torus.
It would be satisfying to show that this coincides with the expression for the Green's function obtained by integrating the heat kernel one gets from the eigenfunction expansion but, as already mentioned, we have not been able to find a corresponding identity in the literature.

%Note that the function
%\be\label{hdef}
%h(z)=-\frac{1}{4\pi} \log\Big[ \sin\frac{\pi z}{2a} \sin\frac{\pi \zb}{2a}\Big]
%\ee
%shares the same properties, except that it is {\it not} periodic in $y=\Im z$ and that the $-\frac{1}{A_0}$-term in the relations analogous to \eqref{Laplg} and \eqref{gisGreen} is absent. 

One can then define the renormalized Green's function at coinciding points ${\cal G}_{\rm R}(z)$ on the torus, after subtracting the short-distance singularity as
\be
{\cal G}_{\rm R} (z)=\lim_{z_1\to z_2\equiv z} \Big( {\cal G}(z_1,z_2) + \frac{1}{4\pi} \log | z_1-z_2 |^2 \Big)
= -\frac{1}{2\pi}\log\Big( \frac{\pi}{a} \ q^{1/6}\prod_{n=0}^\infty (1-q^{2n})^2\Big) \ ,
\ee
with $q=e^{-\pi b/a}$.
As was expected from the isometries of the torus ${\cal G}_R$ is a constant.

%%%%%%%%%%%%%%%%%%%%%%%%%%
\subsection{Green's function and Green's functions at coinciding points on the cylinder}

We now construct the Green's function on the cylinder of length $T$ (coordinate $x$) and circumference $2\pi R$ (coordinate $y$). We choose to impose Neumann boundary conditions at $x=0$ and $x=T$. Let $g$ be the function defined in \eqref{torusg} with $a=T$ and $b=\pi R$:
\be\label{torusgTR}
g(z)= \frac{(\Im z)^2}{8\pi R T}-\frac{1}{4\pi} \log\Big[\t_1\big(\frac{z}{2T}\big\vert i\pi\frac{R}{T}\big) \t_1\big(\frac{\zb}{2T}\big\vert i\pi\frac{R}{T}\big) \Big] +\frac{1}{2\pi} \log \big\vert \eta\big(i\pi\frac{R}{T}\big)\big\vert \ .
\ee
Again, the Neumann boundary conditions are achieved by adding to $g(z_1-z_2)$ the same function with $z_2=x_2+iy_2$ replaced by the appropriate image points. The boundary at $x=0$ requires the image point $z_2^C=-\zb_2=-x_2+iy_2$, while the boundary at $x=T$ would require to add the image points of $z_2$ and $z_2^C$, i.e. $T+(T-x_2+iy_2)=2T+z_2^C$ and $T+(T+x_2+iy_2)=2T+z_2$. However, due to the $2T$-periodicity these points are equivalent to $z_2$ and $z_2^C$ and adding $g$ (or $\wh g$) at these points would result in an over-counting. Of course, this is in agreement with the relation \eqref{cylK} between the heat kernels of the torus and the cylinder, from which the corresponding Green's functions could be obtained by integration over $t$. Thus we let
\be\label{GNcyl}
G^{\rm cyl}(z_1,z_2)={\cal G}(z_1,z_2)+{\cal G}(z_1,-\zb_2)=g(z_1-z_2)+g(z_1+\zb_2) \ .
\ee
Using the symmetry properties \eqref{gsymprop}, one easily verifies that this indeed satisfies Neumann conditions at $x_1=0$ and $T$ as well as at $x_2=0$ and $T$, e.g.
\ba
\del_{n,1} G^{\rm cyl}(z_1,z_2)\big\vert_{x_1=T}&=&g'(T+iy_1-x_2-iy_2) +g'(T+iy_1+x_2-iy_2)\nonumber\\
&=&g'(T-iy_1-x_2+iy_2)-g'(-T-iy_1-x_2+iy_2)=0 \ .
\ea
From \eqref{gisGreen} we see that  $G^{\rm cyl}$ satisfies, for any $x_2\ne 0,T$,
\be\label{LaplGcyl}
\D_{0,z_1} G^{\rm cyl}(z_1,z_2) =\dd^{(2)}(z_1-z_2)-\frac{1}{2\pi R T} \ , 
\ee
where the term $-\frac{1}{2\pi R T}$ arises as $-\frac{2}{A_0^{\rm torus}}$ and equals $-\frac{1}{A_0^{\rm cyl}}$. Integrating the right-hand side of \eqref{LaplGcyl} over the cylinder  then correctly yields 0.

Next, we need to determine the various Green's functions at coinciding points that played an important role for formulating the gravitational action, i.e. $G^{\rm cyl}_\zeta(z)$ and $G^{\rm cyl}_{\rm R,bulk}(z)$. In the present specific case of the cylinder it is useful to first define yet another function $G^{\rm cyl}_{\rm R}(z)$ by
\be\label{GRcyl}
G^{\rm cyl}_{\rm R}(z)=\lim_{z_1\to z_2\equiv z} \left( G^{\rm cyl}(z_1,z_2)
+\frac{1}{4\pi} \log \Big[\sin\frac{\pi(z_1-z_2)}{2T}\sin\frac{\pi(\zb_1-\zb_2)}{2T} \sin\frac{\pi(z_1+\zb_2)}{2T} \sin\frac{\pi(\zb_1+z_2)}{2T}\Big] \right)\ .
\ee
The additional terms subtract the bulk singularity at $z_1\to z_2$, as well as the boundary singularities that occur as $y_1\to y_2$ and $x_1\to x_2\to 0$ or $T$. Explicitly we find that $G^{\rm cyl}_{\rm R}$ only depends on $x=\Re z$ (as well as on $q=e^{-\pi^2 R/T}$, of course):
\be\label{GRcyl2}
G^{\rm cyl}_{\rm R}(z)\equiv G^{\rm cyl}_{\rm R}(x)
=-\frac{1}{2\pi} \log\Big[\frac{\t_1(\frac{x}{T}|i\pi\frac{R}{T})}{\eta(i\pi\frac{R}{T}) \sin\frac{\pi x}{T}}\Big]
-\frac{1}{2\pi} \log\Big[2 q^{1/4} \prod_{n=0}^\infty (1-q^{2n})^2 \Big] \ .
\ee
One sees again, that this is non-singular, even as $x\to 0$ or $x\to T$.
It is clear from its definition that $G^{\rm cyl}_{\rm R}$ satisfies Neumann boundary conditions, as follows also from the explicit expression just given.

However, it is not this quantity $G^{\rm cyl}_{\rm R}$ which enters the gravitational action, but rather $G^{\rm cyl}_\zeta$ or $G^{\rm cyl}_{\rm R, bulk}$. These quantities differ from $G^{\rm cyl}_{\rm R}$ by the following terms:
\ba\label{GzetaGRdiff}
&&\hskip-1.cm\dd G_{\zeta/{\rm R}}^{\rm cyl}(z)  \equiv G^{\rm cyl}_\zeta(z)-G^{\rm cyl}_{\rm R}(z)
=\frac{1}{4\pi} \Big[ 2\log\frac{T\m}{\pi} - E_1(x^2\m^2) - E_1((T-x)^2\m^2) - 2 \log \sin \frac{\pi x}{T}+2\g  \Big] \ ,
\nonumber\\
&&\hskip-1.cm\dd G_{{\rm R,bulk}/{\rm R}}^{\rm cyl}(z) \equiv G^{\rm cyl}_{\rm R,bulk}(z)-G^{\rm cyl}_{\rm R}(z)
= \frac{1}{4\pi} \Big[ 2\log\frac{T\wh\m}{\pi}  - 2 \log \sin \frac{\pi x}{T}+2\g  \Big] \ ,
\ea
The expression $\dd G_{\zeta/{\rm R}}^{\rm cyl}$ is non-singular, even as $x\to 0$ or $x\to T$, as it obviously should be, since all singularities have been removed in the definition of $G_\zeta$, as well as in the one of $G_{\rm R}$.
On the other hand, $\dd G_{\zeta/{\rm R,bulk}}^{\rm cyl}$ is singular as  $x\to 0$ or $x\to T$, and in particular it cannot satisfy the Neumann condition, contrary to $\dd G_{\zeta/{\rm R}}^{\rm cyl}$ which does satisfy the Neumann condition at $x=0,\ T$ for any finite $\m$. Explicitly we have
\be\label{GRbulkcyl}
G^{\rm cyl}_{\rm R,bulk}(x)=-\frac{1}{2\pi} \log \t_1 \big(\frac{x}{T} \big\vert i\pi\frac{R}{T}\big) 
+ \frac{1}{2\pi}\Big(  \log\frac{T\wh\m}{\pi} +\g -  \log\Big[2 q^{1/4} \prod_{n=0}^\infty (1-q^{2n})^2 \Big]  \Big) \ ,
\ee
with only the first term depending on $x$.

%%%%%%%%%%%%%%%%
%\subsection{Laplacian of $G_{\rm R}$ and $G_\zeta$ on the cylinder}

We also need the Laplacian of the Green's function at coinciding points $G_{\rm R}^{\rm cyl}$ and $G_\zeta^{\rm cyl}$.
Recall that $\D_0=-\frac{\del^2}{\del x^2} - \frac{\del^2}{\del y^2} $, so that $\Delta_0 G_{\rm R}^{\rm cyl}(x)=-\frac{\del^2}{\del x^2} G_{\rm R}^{\rm cyl}(x)$ and we find from \eqref{GRcyl2}, using a formula from p 358 of \cite{Erdelyi}, and writing $q=e^{-\pi^2 R/T}$ :
\be\label{LaplGNR}
\Delta_0 G_{\rm R}^{\rm cyl}(x)
%=\left(\frac{1}{2\pi} \log\Big[\frac{\t_1(\frac{x}{T}|i\pi\frac{R}{T})}{\sin\frac{\pi x}{T}}\Big]\right)''
=\frac{1}{2\pi T} \Big( \frac{\t_1'\big(\frac{x}{T}\big\vert i\pi\frac{R}{T}\big)}{\t_1\big(\frac{x}{2T}\big\vert i\pi\frac{R}{T}\big)}-\pi\cot \frac{\pi x}{T}  \Big)'
%=\frac{2}{T} \left(\sum_{n=1}^\infty \frac{q^{2n}}{1-q^{2n}} \sin \frac{2\pi n x}{T} \right)'
=\frac{4\pi}{T^2} \, \sum_{n=1}^\infty \frac{n \, q^{2n}}{1-q^{2n}} \cos \frac{2\pi n x}{T} \ .
\ee
This sum converges for all $x$, as well as all $R\ne 0$ and all finite $T$. 
To get $\D_0  G^{\rm cyl}_\zeta$ we  need to add $\D_0 \dd  G^{\rm cyl}_{\zeta/{\rm R}}$ as obtained from\eqref{GzetaGRdiff}. Using \eqref{Esderiv} we find
\be\label{LapldeltaG}
\D_0\, \dd  G^{\rm cyl}_{\zeta/{\rm R}} (x)
= \frac{1}{2\pi} \big[ \frac{1}{x^2}+2\m^2\big]  e^{-x^2\m^2} + \frac{1}{2\pi} \big[ \frac{1}{(T-x)^2}+2\m^2\big] e^{-(T-x)^2\m^2} - \frac{\pi}{2T^2} \frac{1}{\sin^2\frac{\pi x}{T}}\ .
\ee
Note that this is non-singular for all {\it finite} $\m$, even as $x\to 0$ or $x\to T$. Obviously also, this expression does not have a smooth limit as $\m\to\infty$~: For $x\ne 0,T$, only the last term survives which diverges as $x\to 0,T$. On the other hand, for $x=0$ or $x=T$, one of the first two terms behaves as $\frac{\m^2}{\pi}$ and the limit does not exist. This is in agreement with our general discussion at the end of sect. \ref{zetasec}. However, as explained there, the limit as $\m\to\infty$ of the integrated expression $\int \sqrt{g_0}\, \f\, \D_0 G_\zeta$ should exist, for every $\f$ that obeys Neumann conditions. Now for our cylinder, such a $\f$ can be decomposed on the $\cos  \frac{\pi n x}{T}$ and  one can indeed check very explicitly that $\int_0^T \d x\, \cos \frac{\pi n x}{T} \D_0 \dd  G^{\rm cyl}_{\zeta/{\rm R}} (x)$ is well-defined and has a finite limit as $\m\to\infty$ for every even $n$ (while it vanishes for odd $n$). 
We conclude that
\ba\label{LaplGzetacy2}
\hskip-1.cm\int_0^T \d x\,  \D_0 G_\zeta^{\rm cyl} (x)&=&  \frac{1}{2\pi}\int_0^T \d x\, \f \Big( \big[ \frac{1}{x^2}+2\m^2\big]  e^{-x^2\m^2} + \big[ \frac{1}{(T-x)^2}+2\m^2\big] e^{-(T-x)^2\m^2} 
\nonumber\\
&&\hskip1.5cm - \frac{\pi^2}{T^2} \frac{1}{\sin^2\frac{\pi x}{T}}
+ \frac{8\pi^2}{T^2} \, \sum_{n=1}^\infty \frac{n \, q^{2n}}{1-q^{4n}} \cos \frac{2\pi n x}{T} \Big)
\ , \ \ q=e^{-\pi^2 R/T}  \ ,
\ea
is well-defined and its limit as $\m\to\infty$ exists for every $\f$ obeying the Neumann boundary conditions.

If one thinks of the cylinder as a simple Euclidean version of one compact space and one time dimension, one would like to study the limit where the cylinder becomes infinitely long, i.e. $T\to\infty$.
However, as $\frac{T}{R}\to\infty$, one has $q\to 1$ and the sum over $n$ diverges, hence this expression ceases to be valid. To study the behaviour as $\frac{T}{R}\to\infty$, one must first do the modular transformation $\tau\equiv i\pi\frac{R}{T} \to \wt\tau=-\frac{1}{\tau}=i\frac{T}{\pi R}$. This reads for $\t_1$, as well as for $\t_2$ (which we will need below),
\ba\label{modulartrans}
\t_1(\n|\tau)&=&\frac{i}{(-i\tau)^{1/2}}\exp\big( -i\pi \frac{\n^2}{\tau}\big)\, \t_1\big(\frac{\n}{\tau}\big\vert -\frac{1}{\tau}\big) \ ,
\nonumber\\
\t_2(\n|\tau)&=&\frac{1}{(-i\tau)^{1/2}}\exp\big( -i\pi \frac{\n^2}{\tau}\big)\, \t_4\big(\frac{\n}{\tau}\big\vert -\frac{1}{\tau}\big) \ .
\ea
This will allow us to write the theta functions as sums of powers of $\wt q=e^{i\pi\wt\tau}=e^{-T/R}$. Note that the first argument $\frac{\n}{\tau}$ now is  imaginary which will turn the $\sin$ and $\cos$ in \eqref{LaplGNR} into $\sinh$ and $\cosh$.
We get, through similar manipulations as above,
\ba\label{torusghatTRbis}
\Delta_0 G_{\rm R}^{\rm cyl}(x)&=&\left(\frac{1}{2\pi} \log \frac{\t_1\big(-\frac{ix}{\pi R}\big\vert i\frac{T}{\pi R}\big) }{\sin\frac{\pi x}{T} } -\frac{x^2}{2\pi R T} +\frac{1}{2\pi} \log i\sqrt{\frac{T}{\pi R}} \right)'' 
\nonumber\\
&=&\frac{\pi}{2 T^2}\, \frac{1}{\sin^2\frac{\pi x}{ T}}-\frac{1}{\pi RT} 
-\frac{1}{2\pi R^2}\, \frac{1}{\sinh^2\frac{x}{R}}-
\frac{4}{\pi R^2} \, \sum_{n=1}^\infty \frac{n \, \wt q^{2n}}{1-\wt q^{2n}} \cosh \frac{2n x}{R} \ ,
\ea
where $\wt q=e^{i\pi\wt\tau}=e^{-T/R}$.
%\ba\label{ghat'bis}
%\wh g'(x)&=&\frac{1}{4T} \cot \frac{\pi x}{2T} +\frac{x}{4\pi R T} 
%+\frac{i}{4\pi^2R}\,  \frac{\t_1'\big(-\frac{ix}{2\pi R}\big\vert i\frac{T}{\pi R}\big)}{\t_1\big(-\frac{ix}{2\pi R}\big\vert i\frac{T}{\pi R}\big)}
%\nonumber\\
%&=&\frac{1}{4T} \cot \frac{\pi x}{2T} +\frac{x}{4\pi R T} 
%-\frac{1}{4\pi R} \coth\frac{x}{2R} +\frac{1}{\pi R} \sum_{n=1}^\infty \frac{\wt q^{2n}}{1-\wt q^{2n}}\sinh \frac{n x}{R} \ ,
%\nonumber\\
%\wh g''(x)&=&-\frac{\pi}{8 T^2}\, \frac{1}{\sin^2\frac{\pi x}{2 T}}+\frac{1}{4\pi RT} 
%+\frac{1}{8\pi R^2}\, \frac{1}{\sinh^2\frac{x}{2R}}+
%\frac{1}{\pi R^2} \, \sum_{n=1}^\infty \frac{n \, \wt q^{2n}}{1-\wt q^{2n}} \cosh \frac{n x}{R} \ .
%\ea
Note that the poles at $x=0$ cancel and that this representation as an infinite sum is convergent and finite for all $|x|< T$. Note also that, although not obvious on \eqref{torusghatTRbis},  within this interval $(-T,T)$ these functions are periodic under $x\to x+T$. Then the finiteness at $x=0$ implies finiteness at $x=T$, too.

While \eqref{torusghatTRbis} is a perfectly satisfactory expression, if we think of the $x$-direction as time, we want time to be finite with ``infinite past" and ``infinite future" infinitely far away. Hence we let
\be\label{txshift}
x=\frac{T}{2}+t \quad , \quad t\in[-\frac{T}{2},\frac{T}{2}] \ .
\ee
so that finite $t$ corresponds to values in the ``middle" of the cylinder. It is then natural to first re-express  
the $\t_1\big(\frac{1}{2}+\frac{t}{T}\big\vert i\pi\frac{R}{T}\big)$ appearing in \eqref{LaplGNR}   in terms of $\t_2\big(\frac{t}{T}\big\vert i\pi\frac{R}{T}\big)$ and then use the modular transformation to $\t_4$. This results in
\ba\label{gTplus2t}
\Delta_0 G_{\rm R}^{\rm cyl}(\frac{T}{2}+t)&=&\left(\frac{1}{2\pi} \log \frac{\t_4\big(-\frac{it}{\pi R}\big\vert i\frac{T}{\pi R}\big) }{\cos\frac{\pi t}{T} } -\frac{t^2}{2\pi R T} +\frac{1}{4\pi} \log \frac{T}{\pi R}\right)''
\nonumber\\
%&=&\left(\frac{1}{4T} \tan \frac{\pi t}{T} -\frac{t}{2\pi R T} -\frac{1}{\pi R} \sum_{n=1}^\infty \frac{\wt q^{n}}{1-\wt q^{2n}}\sinh \frac{2n t}{R}\right)' \ ,
%\nonumber\\
&=&-\frac{1}{\pi RT} 
+\frac{\pi}{ 2T^2}\, \frac{1}{\cos^2\frac{\pi t}{T}}
-\frac{4}{\pi R^2} \, \sum_{n=1}^\infty \frac{n \, \wt q^{n} }{1-\wt q^{2n}} \cosh \frac{2 n t}{R}
\ .
\ea
Upon adding $\D_0 G^{\rm cyl}_{\zeta/{\rm R}}$  as given by \eqref{LapldeltaG} with $x=t+\frac{T}{2}$, we also get
\ba\label{LaplGzeta3}
%\begin{array}{|c|}
%\hline\\
 \D_0 G_\zeta^{\rm cyl}(\frac{T}{2}+t)
&=&
-\frac{1}{\pi RT} 
+\frac{1}{2\pi}\big[\frac{1}{(\frac{T}{2}+t)^2}+2\m^2\big] e^{-(\frac{T}{2}+t)^2\m^2}  
+\frac{1}{2\pi}\big[\frac{1}{(\frac{T}{2}-t)^2}+2\m^2\big] e^{-(\frac{T}{2}-t)^2\m^2}  
%\nonumber\\
\nonumber\\
&&-\frac{4}{\pi R^2} \, \sum_{n=1}^\infty \frac{n \, \wt q^{n} }{1-\wt q^{2n}} \cosh \frac{2 n t}{R}
\ .
%\\
%\\
%\hline
%\end{array}
\ea
Again, we have a finite and convergent expression for all $|t|<\frac{T}{2}$. On the other hand, as $t\to \pm \frac{T}{2}$, both the $\frac{1}{2\pi(\frac{T}{2}\pm t)^2}  $ and the sum diverge. However, we know  that $\Delta_0 G_{\rm R}^{\rm cyl}$ and $\D_0 G_\zeta^{\rm cyl}$ are finite for all $t\in[-\frac{T}{2},\frac{T}{2}]$ (as long as $\m$ is finite).

It is now easy to see what happens for finite $t$ and $T\to\infty$. In this limit $\wt q=e^{-T/R}\to 0$  and the sum does not contribute. Also, the factors $e^{-(\frac{T}{2}\pm t)^2\m^2}$ vanish exponentially for all finite $\m$.  Thus
\be\label{LaplGNRtlimit}
\Delta_0 G_\zeta^{\rm cyl}(\frac{T}{2}+t)\simeq -\frac{1}{\pi RT}+{\cal O}\big(\m^2 e^{-T^2\m^2}, e^{-T/R}\big)
= -\frac{2}{A_0^{\rm cyl}} +{\cal O}\big(\m^2 e^{-T^2\m^2}, e^{-T/R}\big)
\ , \quad \text{$t$ finite and $T\to\infty$} \ .
\ee

\subsection{The gravitational action on the cylinder}

We are now in position to explicitly give the gravitational action on the cylinder. The Liouville part has a universal form, while the purely cosmological term is not of much interest since it gets combined with the corresponding counterterm. Hence, we will only display the genuine order $m^2 A$-term that generalises the Mabuchi (and Aubin-Yau) actions. From \eqref{Sgrav5} we get
\be\label{Sgravcyl1}
S_{\rm grav}^{\rm cyl}[g,g_0]\Big\vert_{m^2 A{\rm -term}}
= \frac{m^2 A}{4} \,  \int_0^T \d x \int_0^{2\pi R} \d y\, \Big(-\frac{1}{2}\f\D_0\f  +\frac{1}{\pi A} \s e^{2\s} 
+\frac{1}{2\pi} \D_0\f\,  \log \t_1 \big(\frac{x}{T} \big\vert i\pi\frac{R}{T}\big)  \Big) \ ,
\ee
with\footnote{Since $\log\t_1$ does not depend on $y$, the $y$-derivative is a total derivative and gives a vanishing result.}
$\D_0=-\frac{\del^2}{\del x^2}-\frac{\del^2}{\del y^2}$. 
Alternatively, we can use \eqref{Sgrav7} and obtain
\ba\label{Sgravcyl2}
\hskip-1.cm S_{\rm grav}^{\rm cyl}[g,g_0]\Big\vert_{m^2 A{\rm -term}}
&=& \frac{m^2 A}{4} \, \lim_{\m\to\infty} \int_0^T \d x \int_0^{2\pi R} \d y\, \Bigg\{-\frac{1}{2}\f\D_0\f  +\frac{1}{\pi A} \s e^{2\s} 
\nonumber\\
&&+\f\, \Bigg[
\frac{\pi}{2T^2} \frac{1}{\sin^2\frac{\pi x}{T}}
- \frac{4\pi}{T^2} \, \sum_{n=1}^\infty \frac{n \, e^{-2n\pi^2 R/T}}{1-q^{-4n\pi^2 R/T}} \cos \frac{2\pi n x}{T}
\nonumber\\
&&\hskip8.mm -\frac{1}{2\pi} \big[ \frac{1}{x^2}+2\m^2\big]  e^{-x^2\m^2} - \frac{1}{2\pi} \big[ \frac{1}{(T-x)^2}+2\m^2\big] e^{-(T-x)^2\m^2} 
\Bigg] \Bigg\} .
\ea
As mentioned above, with $\f$ obeying Neumann boundary conditions, the limit $\m\to\infty$ indeed exists (and equals \eqref{Sgravcyl1}, of course).

In order to study the limit of an infinitely long cylinder, we have seen that one has to set $x=\frac{T}{2}+t$ and use \eqref{LaplGzeta3} in order to obtain the limit $T\to\infty$ as given by  \eqref{LaplGNRtlimit}. Thus
\be\label{Sgravcyl3}
S_{\rm grav}^{\infty \ \rm cyl}[g,g_0]\Big\vert_{m^2 A{\rm -term}}
= \lim_{T\to\infty}\, \frac{m^2 A}{4} \,
\int_{-T/2}^{T/2} \d t \int_0^{2\pi R} \hskip-2.mm \d y\  \Bigg\{-\frac{1}{2}\f\D_0\f  +\frac{1}{\pi A} \s e^{2\s} 
+\frac{2}{A_0} \f \Bigg\} \ .
\ee
Of course, $A_0=2\pi R T$ and $A$ also go to infinity in this limit, and the second and third terms of the Lagrangian have finite coefficients. As for the kinetic term, only those eigenvalues of $\D_0$ that scale as $\frac{1}{A}$ give finite contributions.

Comparing \eqref{Sgravcyl3} with the Mabuchi action of a manifold without boundary as defined in \eqref{Mab1}, we see that this corresponds to the Mabuchi action with $h=0$ and vanishing background curvature (i.e. $R_0=0$):
\be\label{infcylMab}
S_{\rm grav}^{\infty \ \rm cyl}[g,g_0]\Big\vert_{m^2 A{\rm -term}}
= \frac{m^2 A}{16\pi} S_{\rm M}[g,g_0]\Big\vert_{h=0,\ R_0=0} \ .
\ee
Of course, this equality is to be understood as an equality of the Lagrangian densities, rather than of the actions.
While $R_0=0$ was to be expected for a cylinder, the replacement $\chi=2(1-h)\to 2$ was, maybe, not that obvious to guess.

\vskip5.mm
\noindent
{\Large\bf Acknowledgements}

\vskip5.mm
\noindent
We are grateful to the referee for suggesting various improvements of the manuscript.

\vskip8.mm

%\newpage
\noindent
{\Large\bf Appendix}

%%%%%%%%%%%%%%%%%%%%%%%%%%%%%%%%%%%%%
\begin{appendix}
\section{Boundary integrals for two-dimensional manifolds}

Integration by parts on a two-dimensional manifold with boundaries generates boundary terms. The natural way to implement this is via Stoke's theorem for the integration of an exact 2-form. Let's work this out.

Let $\a=\a_a \d x^a$ be a 1-form on $\cM$. Let $\wh\a$ be its restriction to $\del\cM$ (i.e. the pullback of $\a$ under the inclusion map of $\del\cM$ into $\cM$):  $\wh\a=\a_a \d \wh x^a$ where the $\d\wh x^a$ are the ``projections" of the $\d x^a$ on the tangent to $\del\cM$. We define the not necessarily normalized tangent vector $t^a$ as $\d\wh x^a=t^a \d l$ where $\d l$ is the proper length one-form on $\del\cM$. Thus 
\be
\wh\a=\a_a t^a \d l \ . 
\ee
As an example,
let $\cM$ be the unit sphere $S^2$, with standard coordinates $\t,\vf$, with the polar cap $\t\le \t_0$ removed. Then $\del\cM$ is the circle at $\t=\t_0$, so that $\d\wh\t=0$ and $\d\wh\vf=\d\vf$. Now $\d l=\sin\t\, \d\vf$ so that $t^\t=0$ and $t^\vf=\frac{1}{\sin\t}$.
Integrals of the 1-form $\wh\a$ over $\del\cM$ can be immediately evaluated as
\be
\int_{\del\cM}\wh\a=\int_{\del\cM} \a_a \d\wh x^a=\int_{\del\cM} \a_a t^a \, \d l\ ,
\ee
without the need to introduce a metric. However,
sometimes, the 1-form $\a$ is the Hodge dual of some other 1-form $\b$, i.e. $\a={}^*\b$, and this requies a metric. Indeed, we have 
\be
{}^*\d x^a=g^{ab}\e_{bc}\d x^c=g^{ab} \sqrt{g}\, \wh\e_{bc}\d x^c\ ,
\ee
where $\wh\e_{12}=1,\ \wh\e_{21}=-1$. Then ${}^*\b=\b_a{}^*\d x^a=\b_a g^{ab} \e_{bc}\d x^c$ and $\wh{{}^*\b}=\b_a g^{ab}\e_{bc}\d \wh x^c =\b_a g^{ab} \e_{bc} t^c \d l$. One then defines the not necessarily normalized normal vector $n^a$ as
\be
n^a=g^{ab} \e_{bc} t^c=g^{ab}\sqrt{g}\, \wh e_{bc} t^c 
\quad \Rightarrow\quad
\wh{{}^*\b}=\b_a n^a \d l \ .
\ee
Note that $g_{ab} n^a t^b=\e_{ab}t^a t^b=0$, as expected. For the above example of the sphere with the polar cap removed we have
$g^{\vf\vf}=\frac{1}{\sin^2\t},\ \sqrt{g}=\sin\t$ and $n^\t=\sin\t_0,\ t^\vf=1,\ n^\vf=0$.

Integration by parts follows from Stoke's theorem, 
\be\label{Stokes}
\int_{\cM} \d \g=\int_{\del\cM} \wh\g \ ,
\ee 
for any one-form $\g$. We are mostly interested in $\d\g$ being a kinetic term, i.e. $\g=\f\, {}^* \d\f={}^* ( \f\,\d\f)$. Then $\d\g=\d\f\wedge {}^*\d\f+\f \, \d\, {}^*\d\f$. We have
$\d\f\wedge {}^*\d\f
=\d x^1\wedge \d x^2\, \sqrt{g}\,  g^{ab} \del_a\f \del_b\f$, as well as
$\f \, \d{}^*\d\f=\f\, \del_a (g^{ab}\sqrt{g} \,\del_b\f) \d x^1\wedge\d x^2=-\d x^1\wedge\d x^2\sqrt{g}\, \f\D\f$, 
where, as throughout this paper,  have defined the scalar Laplace operator with a minus sign so that its eigenvalues are positive. Then Stoke's theorem gives the following integration by parts formula
\be
\int_{\cM} \d^2 x \sqrt{g}\, g^{ab} \del_a\f \del_b\f
=\int_{\cM}\d^2 x \sqrt{g}\, \f \D \f
+\int_{\del\cM} \d l\, n^a\, \f\, \del_a \f \ .
\ee

%%%%%%%%%%%%%%%%%%%%%%%%%%%%%%%%%%%%%
\section{$E_s(x)$ and the incomplete $\G$-function}

The function $E_s(x)$  is defined as 
\be\label{Esdef}
E_s(x)=\int_1^\infty \d u\, u^{-s}\, e^{-x u}  \ , \ x>0 \ .
\ee
By the change of variables $t=xu$ ($x>0$) it can be related to the incomplete $\G$-function:
\be\label{gammadef}
\G(a,x)=\int_x^{\infty} \d t\, t^{a-1}\, e^{-t}
\quad , \quad
E_s(x)=x^{s-1}\, \G(1-s,x) \ , \quad \ x>0 \ .
\ee
Note that for $\Re s>1$ the definition of $E_s(x)$ continues to make sense for $x=0$, but the previous relation doesn't. It follows from the definition \eqref{Esdef} that
\be\label{Esderiv}
\frac{\d}{\d x} E_s(x)=-E_{s-1}(x) \ ,
\quad , \quad
\frac{\d}{\d x} E_1(x)=-E_0(x)=-\frac{e^{-x}}{x} \ ,
\ee

We are interested in the behaviour of $E_s(x)$ for $s$ in the vicinity of 1 and small $x$, i.e. for $0<x\ll1$. We let $s=1-\e$ such that $E_{1-\e}(x)= x^{-\e}\, \G(\e,x)$. As is clear from its definition, $\G(\e,x)$ is singular if both $\e$ and $x$ go to 0. However, for $x>0$, $\G(a,x)$ is well-defined for all $a\in {\bf C}$. In particular, $\G(0,1)$ is a finite number. We have 
\be\label{Gammaeta}
\Gamma(\e,x)=\G(\e,1)+\int_x^1\d t\, t^{-1+\e}e^{-t}= \G(\e,1)+\sum_{n=0}^\infty \frac{(-)^n}{n!} \int_x^1\d t\, t^{-1+\e+n}
%\nonumber\\ &=&
=\G(\e,1)+\sum_{n=0}^\infty \frac{(-)^n}{n!}\frac{1-x^{n+\e}}{n+\e} \ .
\ee
Only the $n=0$ term is potentially divergent as $\e\to 0$ and $x\to 0$, and separating it from the rest of the sum we get:
\be\label{Gammaeta2}
\Gamma(\e,x)=\frac{1-x^\e}{\e} +\Big[\G(\e,1)+\sum_{n=1}^\infty \frac{(-)^n}{n! (n+\e)} \Big]-  \sum_{n=1}^\infty \frac{(-x)^n}{n! (n+\e)} \ ,
%\nonumber\\ &=&
\frac{1-x^\e}{\e} -\g -  \sum_{n=1}^\infty \frac{(-x)^n}{n! \, n} +{\cal O}(\e) \ ,
\ee
where we used $\G(0,1)+\sum_{n=1}^\infty \frac{(-)^n}{n! \ n}=-\g$.
Similarly,
\be\label{Esx}
E_s(x)=\frac{1}{s-1}-x^{s-1}\Big[ \frac{1}{s-1} +\g +  \sum_{n=1}^\infty \frac{(-x)^n}{n! \, n} +{\cal O}(s-1) \Big]
\ .
\ee
If we first let $\e=1-s\to 0$ we get
\be\label{gammaeps0}
E_1(x)=\G(0,x)=-\log x -\g  -  \sum_{n=1}^\infty \frac{(-x)^n}{n! \, n}\ ,
\ee
while if we first let $x\to 0$ we get
\be\label{gammax0}
%\G(\e,0)=\frac{1}{\e} -\g +{\cal O}(\e) \quad , \quad
E_s(0)=  \frac{1}{s-1} \ , \ s>1 \ .
\ee
Using these formulae, it is straighforward to show that
\be\label{EsEsprimelimit}
\lim_{s\to 1} \Big[E_s(x) + (s-1)\frac{\d}{\d s} E_s(x)\Big] = E_1(x) \quad , \quad x>0 \ .
\ee
and
\be\label{EsEsprimelimitx=0} 
E_s(0) + (s-1)\frac{\d}{\d s} E_s(0)= 0  \quad , \quad s>1 \ .
\ee
This relation holds exactly for all $s>1$ and, hence, also in the limit $s\to 1$. Not only the limit 
as $x\to 0$ of 
\eqref{EsEsprimelimit} is different from \eqref{EsEsprimelimitx=0}, it is also singular.
 This non-commutativity of the limits $s\to 1$ and $x\to 0$ can be traced back to the behaviour of 
 $\frac{\d}{\d s} x^{s-1}= x^{s-1}\log x$. If we first let $x\to 0$ assuming $s>1$ it yields 0, while letting first $s\to 1$ assuming $x>0$ we get $\log x$.

%%%%%%%%%%%%%%%%%%%%%%%%%%%%%%%%%%%%%
\section{Some additional variational formulae}

When establishing the gravitational action,  we had chosen to study the properties of $\wt G_{\rm R,bulk}$ rather than those of $\wt G_\zeta$. If one chooses to study the properties of the latter instead, one needs, in particular, the variation of $\wt G^{(0)}_\zeta$ for finite $\m$.  This requires the use of some additional variational formulae which we summarize in this appendix.

First, the variation of $\wt G_\zeta$ also involves the variation of $E_1\big(\frac{\ell^2(y,y_C^i) \m^2}{4}\big)$.
For $x\to y$ and close to the boundary one has
\be\label{ellbounvar}
\dd \ell^2(y,y_C^i) \simeq 2 \dd\s(y_B) \, \ell^2(y,y_C^i)  \ .
\ee
It follows that
\be\label{deltaE}
\dd E_1\big(\frac{\ell^2(y,y_C^i) \m^2}{4}\big)=E_1' \big(\frac{\ell^2(y,y_C^i) \m^2}{4}\big) \frac{\ell^2(y,y_C^i) \m^2 }{2}\, \dd\s(y_B^i) = -2 e^{-\frac{\ell^2(y,y_C^i) \m^2}{4}} \dd\s(y_B^i) \ ,
\ee
where we used \eqref{Esderiv} Thus, we get
\be\label{Gzetavar}
\dd G_\zeta(x)=-2 m^2 \int \d^2 z \sqrt{g} \big( G(x,z)\big)^2 \dd\s(z) + \frac{\dd\s(x)}{2\pi}  + \sum_{\del\cM_i} e^{-\frac{\ell^2(x,x_C^i) \m^2}{4}} \frac{\dd\s(x_B^i)}{2\pi}\ .
\ee
One also encounters $\sum_i \int_{\del\cM_i} \d l \, \dd\s$~:
\be\label{lenghtvar}
\sum_i \int_{\del\cM_i} \d l \, \dd\s(x_B)=\sum_i \dd \int_{\del\cM_i} \d l =\dd L(\del\cM) \ ,
\ee
where $L(\del\cM)$ is the total length of the boundary. The finite variation of $\wt G^{(0)}_\zeta$ follows from \eqref{Gzetavar} in the zero-mass limit as
\be\label{Gzetavarfinite}
\wt G^{(0)}_\zeta(x,g)-\wt G^{(0)}_\zeta(x,g_0)=
\f(x) +\frac{\s(x)}{2\pi}-S_{AY}[g_0,g] -\frac{1}{4\pi}   \sum_{\del\cM_i} 
\Big( E_1\big(\frac{\ell^2_g(x,x_C^i) \m^2}{4} \big) - E_1\big(\frac{\ell^2_{g_0}(x,x_C^i) \m^2}{4} \big)\Big)\ ,
\ee
Finally, at finite $\m$ the gravitational action can then be found to be
\ba\label{Sgrav5bis}
S_{\rm grav}[g,g_0]&\hskip-2.mm=\hskip-2.mm &-\frac{1}{24\pi}  S_{L}[g,g_0] +\frac{1}{2}\log\frac{A}{A_0}
%\nonumber\\ &&
+\frac{ m^2\, (A-A_0)}{2 A_0}  \Big( \Phi_G[g_0]+\frac{1}{4\sqrt{\pi}\, \m} L(\del\cM,g_0) \Big)
\nonumber\\
&&+ \frac{m^2 A}{4} \, \Bigg[ \int\sqrt{g_0}\Big(-\frac{1}{2}\f\D_0\f  +\frac{1}{\pi A} \s e^{2\s} -\f\,  \D_0\wt G^{(0)}_\zeta[g_0] \Big) -\frac{1}{4\sqrt{\pi}\m} \int_{\del\cM} \d l_0\, \D_0\f\ + {\cal O}\big(\frac{1}{\m^2}\big)  \Bigg]
\nonumber\\
&&+ {\cal O}(m^4)\ .
\ea

\end{appendix}

%%%%%%%%%%%%%%%%%%%%%%%%%%%%%%%%%%%%%%%%%%%%

%%%%%%%%%%%%%%%%%%%%%%%%%%%%%%%%%%%%%%%%%%%%

%%%%%%%%%%%%%%%%%%%%%%%%%%%%%%%%%%%%%%%%%%%%

%%%%%%%%%%%%%%%%%%%%%%%%%%%%%%%%%%%%%%%%%%%%

%%%%%%%%%%%%%%%%%%%%%%%%%%%%%%%%%%%%%%%%%%%%

%%%%%%%%%%%%%%%%%%%%%%%%%%%%%%%%%%%%%%%%%%%%

%\vskip5.mm
%\noindent
%{\Large\bf Acknowledgements}

\vskip3.mm
%\noindent
%C.D. is supported by a fellowship from ......
%\vskip5.mm

%\newpage

%%%%%%%%%%%%%%%%%%%%%%%%%%%%%%%%%%%%%%%%%%%%%%%%%%%%%%%%%%%%

%%%%%%%%%%%%%%%%%%%%%%%%%%%%%%%%%%%%%%%%%%%%
	
\end{document}